\documentclass[12pt]{iopart}
\pdfoutput=1
\usepackage{graphicx}
\expandafter\let\csname equation*\endcsname\relax
\expandafter\let\csname endequation*\endcsname\relax
\usepackage{amsmath}
\usepackage{amssymb}
\usepackage{bm}
\usepackage{cancel}
\usepackage[normalem]{ulem}
\usepackage{subfig}
\usepackage{placeins}
\usepackage{color}
\usepackage{iopams}
\usepackage{cite}

\usepackage{hyperref}
\newcommand{\deltaf}{\delta\! f}
\newcommand{\st}{\text{st}}

\begin{document}

\title[Pulling speed dependence of the
  unfolding pathway of proteins]{Understanding the dependence on the
pulling speed of the unfolding pathway of proteins}
\author{C.~A.~Plata$^{1}$, F. Cecconi$^{2}$, M.~Chinappi$^{3}$,  and A.~Prados$^{1}$}

\address{$^{1}$F\'{\i}sica Te\'{o}rica, Universidad de Sevilla,
Apartado de Correos 1065, E-41080 Seville, Spain, EU}
\address{$^{2}$CNR-Istituto dei Sistemi Complessi (ISC), Via dei
  Taurini 19, I-00185 Rome, Italy, EU}
\address{$^{3}$Center for Life Nano Science,
Istituto Italiano di Tecnologia (IIT), Via Regina Elena 291, I-00161, Rome, Italy, EU}

\begin{abstract}

  The dependence of the unfolding pathway of proteins on the pulling
  speed is investigated. This is done by introducing a simple
  one-dimensional chain comprising $N$ units, with different
  characteristic bistable free energies. These units represent either
  each of the modules in a modular protein or each of the intermediate
  ``unfoldons'' in a protein domain, which can be either folded or
  unfolded. The system is pulled by applying a force to the last unit
  of the chain, and the units unravel following a preferred
  sequence. We show that the unfolding sequence strongly depends
    on the pulling velocity $v_{p}$. In the simplest situation, there
    appears a critical pulling speed $v_{c}$: for pulling speeds
    $v_{p}<v_{c}$, the weakest unit unfolds first, whereas for
    $v_{p}>v_{c}$ it is the pulled unit that unfolds first. By means
    of a perturbative expansion, we find quite an accurate expression
    for this critical velocity.

\end{abstract}




\maketitle

\section{\label{sec:intro} Introduction}

Since the late twentieth century, research on the mechanical stability
of macromolecules turned a prolific field due to the advances in
manipulation techniques of individual biomolecules, usually termed
single-molecule experiments. One of the most important techniques is
atomic force microscopy (AFM), in which a biomolecule is stretched
between a rigid platform and the tip of the cantilever
\cite{R06,KyL10,MyD12,HyD12}. In these experiments, the
controlled parameter is either the length of the macromolecule
(length-control protocols) or the force exerted over it (force-control
protocols), and its conjugated magnitude is measured. As a result, a
force-extension curve (FEC) is obtained, which characterizes the
elasto-mechanical behaviour of the macromolecule and provides
fundamental information about its unfolding pathway
\cite{SCyC96,LyK99,MyR05,CKyL08,COFMBCyF99,FMyF00,LOSTyB01,CByB03,KyT00,HDyT06}.

In a typical pulling experiment, the end-to-end distance of the
  molecule $L$ is increased with a pulling rate $v_{p}$, that is,
  $dL(t)/dt=v_{p}$. Remarkably, the FEC exhibits a sawtooth pattern
\cite{COFMBCyF99,FMyF00,LOSTyB01,CByB03,LDSTyB02,HDyT06} showing how the
macromolecule comprises several structural units or blocks, in general
each one with different stability properties. Each block unfolds
individually causing a drop in the measured force. The unfolding
pathway is, basically, the order and the way in which the structural
blocks of the macromolecule unravel.

Recent studies show that the pulling velocity plays a relevant role in
determining the unfolding pathway
\cite{HDyT06,LyK09,GMTCyC14,KHLyK13}. Different unfolding pathways are
observed depending on (a) which of the ends (C-terminus or N-terminus)
the domain is actually pulled from and (b) the pulling speed. It has
been claimed that it is the inhomogeneity in the distribution of the
force across the protein, for high pulling speeds, that causes the
unfolding pathway to change
\cite{HDyT06,LyK09,GMTCyC14,KHLyK13}. Nevertheless, to the best of our
knowledge, a theory that explains this crossover is still
lacking.

We focus on the unfolding pathway of protein domains composed of
several stable structural units \cite{LyK09,GMTCyC14,ByR08}.  In this
respect, a good candidate is represented by the Maltose Binding
Protein (MBP), a stable and well characterized protein comprising two
domains. MBP has been recently employed in studies on mechanical
unfolding and translocation
\cite{GMTCyC14,ByR08,bacci2012role,bacci2013protein,merstorf2012wild,
  aggarwal2011ligand,kotamarthi2014mechanical}.  In particular, Bertz
and Rief \cite{ByR08} identified four intermediate states in the FEC
of mechanical denaturation experiments, each one associated to the
unravelling of a specific unit. The authors termed such units
``unfoldons'' and determined their typical unfolding
sequence. However, simulations \cite{GMTCyC14} showed that this
unfolding scenario holds only for low pulling speed as the pathway
depends on the rate at which the molecule is pulled: at very low pulling rates,
it is the weakest unfoldon that unfolds
first, while at higher rates the first unfoldon to unravel is the
pulled one.

The main purpose of this paper is to develop
a physical theory that predicts the observed
dependence of the unfolding pathway on the pulling
velocity. Specifically, we do
so in the adiabatic limit, that is, the regime of slow pulling
speeds that allow the system to sweep the whole stationary branches
of the FEC \cite{PCyB13,BCyP15}. Also, we would like to stress that,
although our approach is applied to the unfolding of a single protein,
it extends to the unfolding of polyproteins comprising several
domains (modular proteins), which unfold sequentially \cite{FMyF00}.

The plan of the paper is as follows. In section \ref{sec:model}, we
put forward our model to investigate the unfolding pathway of
proteins. In general, the free energy characterizing each unit
(unfoldon or module) of the protein is different, that is, there is a
certain degree of asymmetry (or disorder) in the free energies.
Moreover, we discuss the role of thermal noise and the
  (ir)relevance of the details of the device controlling the length of
  the protein. Section \ref{sec:real_chain} is devoted to the
analysis of the pulling of the model, by means of a perturbative
expansion in both the asymmetry of the free energies and the pulling
speed. In sections \ref{sec:asymmetry} and \ref{sec:pulling_speed} we
obtain the corrections introduced by the asymmetry and the finite
value of the pulling speed, respectively.  In section
\ref{sec:critical_speed} we show that, in the simplest situation,
there naturally appears a critical pulling speed $v_{c}$, below
(above) which it is the weakest (pulled) unit that unfolds first.  In
general, when more than one unit has a different free energy, we show
that for low (high) enough pulling velocity, it is still the weakest
(pulled) unit that unfolds first, but other pathways are present for
intermediate velocities.  Numerical results for some particular
situations are shown in section \ref{sec:numerics}. They are compared
to the analytical results previously derived, and a quite good
agreement is found. Finally, section \ref{sec:conclusions} deals with
the main conclusions of the paper. The appendices cover some technical
details that we have omitted in the main text.

\section{\label{sec:model} The model}
\begin{figure}
\centering
  \includegraphics[width=0.85 \textwidth]{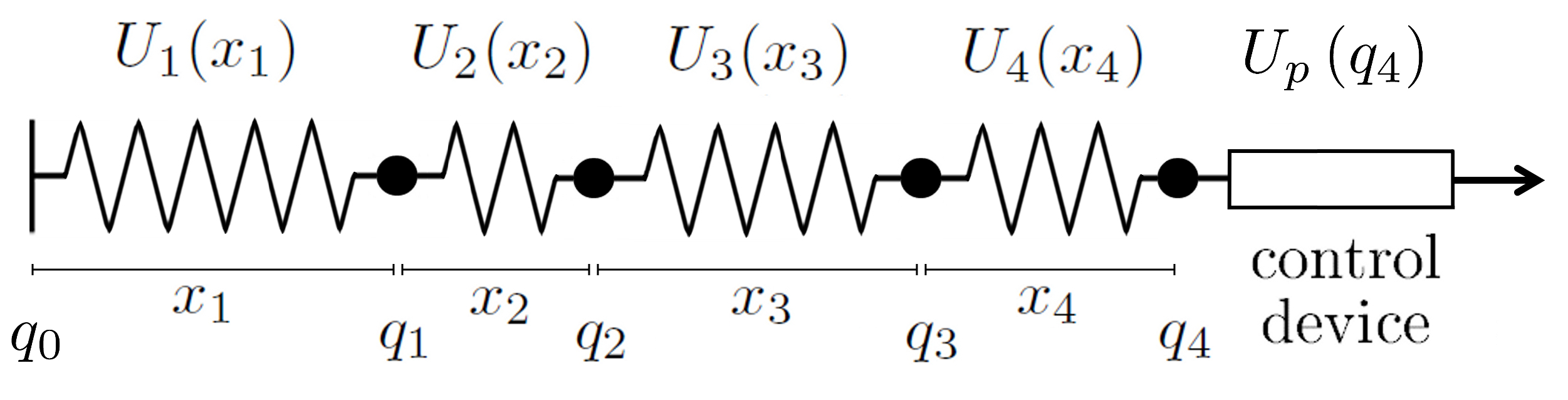}
\caption{\label{fig:1}
   Sketch of the model for a protein with four units. Each unit
     is represented by a nonlinear spring with potential $U_i(x_i)$,
     in which  $x_i$ is the unit's extension. The beads mark the
     coordinates $q_{i}$ of their endpoints, so that
     $x_{i}=q_{i}-q_{i-1}$ (by definition, $q_{0}=0$). Finally, the
     rectangle stands for the device attached to the pulled end $q_4$,
     which controls either the force applied to the molecule
     (force-control) or its end-to-end distance (length-control). The
     contribution of this device to the system's free energy is
     $U_p(q_4)$, as shown by \eqref{eq:4} and \eqref{eq:6b}.
}
\end{figure}

Let us consider a certain protein domain comprising $N$ unfoldons (or
a polyprotein composed of $N$, possibly different, modules). From now
on, we will refer to these unfoldons or modules as units.  When the
molecule is submitted to an external force $F$, the simplest
description is to portray it as a one-dimensional chain, where the
end-to-end extension of the $i$-th unit in the direction of the force
is denoted by $x_{i}$.  In a real AFM experiment, the molecule is
attached as a whole to the AFM device and stretched.  Following
Guardiani et al. \cite{GMTCyC14}, we model this system with a sequence
of nonlinear bonds, as in figure \ref{fig:1}: the endpoints of
  the $i$-th unit are denoted by $q_{i-1}$ and $q_{i}$, so that its
  extension $x_{i}$ is
\begin{equation}
  \label{eq:3}
  x_{i}=q_{i}-q_{i-1}, \quad i=1,\ldots,N.
\end{equation}
The evolution of the system follows the coupled overdamped
Langevin equations
\begin{equation}
  \label{eq:5}
  \gamma \dot{q}_{i}=-\frac{\partial}{\partial q_{i}} G(q_{1},\ldots,q_{N})+\eta_{i},
\end{equation}
in which the $\eta_{i}$ are the Gaussian white noise terms, such that
$\langle \eta_{i}(t)\rangle=0$ and
$\langle \eta_{i}(t) \eta_{j}(t')\rangle=2 \gamma k_{B} T \delta_{ij}
\delta(t-t')$,
with $\gamma$ and $T$ being the friction coefficient of each unit (the
same for all) and the temperature of the fluid in which the protein is
immersed, respectively ($k_B$ is the Boltzmann constant).  The global
free energy function of the system is
\begin{equation}
  \label{eq:4}
  G(q_{1},\ldots,q_{N})=\sum_{i=1}^{N} U_{i}(q_{i}-q_{i-1})+U_{p}(q_{N}) \, .
\end{equation}
In \eqref{eq:4}, $U_{p}(q_{N})$ is the contribution to the free
energy introduced by the force-control or length-control device (see
below), while $U_i(x_i)$ is the single unit contribution to $G$.

The total length of the system is given by
\begin{equation}
  \label{eq:6}
  \sum_{i=1}^{N} x_{i}=q_{N}.
\end{equation}
In force-control experiments, the applied force $F$ is a given
function of time, whereas in length-control experiments the device
tries to keep the total length $q_{N}$ equal to the desired value $L$,
also a certain function of time.
The corresponding contributions to the free energy are
\begin{subequations}\label{eq:6b}
\begin{equation}
  U_{p}(q_{N})=-F q_{N}, \qquad \text{force-control},
\end{equation}
\begin{equation}
  U_{p}(q_{N})=\frac{1}{2}k_{p}(q_{N}-L)^{2}, \qquad \text{length-control},
\end{equation}
\end{subequations}
in which $k_{p}$ stands for the stiffness of the length-control
device. The length is perfectly controlled in the limit $k_{p}\to\infty$,
when $q_{N}=L$ for all times. A sketch
of the model is presented in figure \ref{fig:1}.

An apparently similar system, in which each module of the chain
follows the Langevin equation
$\gamma \dot{x}_{i}=-\partial G/\partial x_{i}+\eta_{i}$ has been
recently analyzed \cite{BCyP14,BCyP15}. In this approach, the modules
are completely independent in force-controlled experiments because
these Langevin equations completely neglect the spatial structure of
the chain. While this simplifying assumption poses no problem for the
characterization of the force-extension curves in \cite{BCyP15}, it is
not suited for the investigation of the unfolding pathway, in which
the spatial structure plays an essential role. The spatial
  structure of biomolecules can be described in quite a realistic way
  by using a model proposed by Hummer and Szabo several years ago to
  investigate their stretching \cite{HyS03}, but the simplified
  picture which follows from figure \ref{fig:1} makes an analytical
  approach feasible.

Now, we turn to look into the unfolding pathway of this system. As the
evolution equations are stochastic, this pathway may vary from one
trajectory of the dynamics to another.  Nevertheless, in many
experiments \cite{LyK09,GMTCyC14,ByR08} it is observed that there is a
quite well-defined pathway, which suggests that thermal fluctuations
do not play an important role in determining it. Physically, this
  means that the free energy barrier separating the unfolded and
  folded conformations at coexistence (that is, at the critical force,
  see below) is expected to be much larger than the typical energy
  $k_{B}T$ for thermal fluctuations. Therefore, we expect the thermal
noise terms in our Langevin equations to be negligible and,
consequently, they will be dropped in the remainder of the
  theoretical approach developed in the paper. Of course, if the
  unfolding barrier for a given biomolecule were only a few $k_{B}T$s,
  the thermal noise terms in the Langevin equations could not be
  neglected and our theoretical approach would have to be changed.

In order to undertake a theoretical analysis of the stretching
dynamics, one further simplification of the problem will be
introduced. We consider that the device controlling the length is
perfectly stiff, and the total length $q_{N} = L$ does not fluctuate
for all times.  We expect this assumption to have little impact on the
unfolding pathway: otherwise, the latter would be more a property of
the length-control device than of the chain.  In fact, we  show in
section \ref{sec:numerics} that the unfolding order is not affected by
this simplification.  For perfect length-control, the mathematical
problem is identical to that of the force-control situation, but now
the force $F$ is an unknown (Lagrange multiplier) that must be
calculated at the end by imposing the constraint
$q_{N}=\sum_{i}x_{i}=L$.  Therefore, the extensions $x_i$'s obey the
deterministic equations
\begin{subequations}
\label{eq:7}
\begin{eqnarray}
  \gamma\dot{x}_{1}  =  -U'_{1}(x_{1})+U'_{2}(x_{2}), \\
  \gamma\dot{x}_{i}  =  -2U'_{i}(x_{i})+U'_{i+1}(x_{i+1})+U'_{i-1}(x_{i-1}),
     \quad 1<i<N, \\
  \gamma\dot{x}_{N}  =  -2U'_{N}(x_{N})+U'_{N-1}(x_{N-1})+F, \\
  F =\gamma v_{p}+U'_{N}(x_{N}).
\end{eqnarray}
\end{subequations}
We have introduced the pulling speed
\begin{equation}
  \label{eq:8}
  v_{p}\equiv \dot{L},
\end{equation}
which is usually time independent.

We assume that $U_i(x_i)$ allows for bistability in a certain range of
the external force $F$, in the sense that $U_{i}(x_{i})$
$-F x_i$ is a double-well potential with two minima, see figure
\ref{fig:2}. Therefore, in that force range, each unit may be
either folded, if $x_{i}$ is in the well corresponding to the minimum
with the smallest extension, or unfolded, when $x_{i}$ belongs to the
well with the largest extension.
If the length is kept constant ($v_{p}=0$), there is an equilibrium
solution of \eqref{eq:7},
\begin{equation}
  \label{eq:17}
  U'_{1}(x_{1}^{\st})=U'_{2}(x_{2}^{\st})=\cdots=U'_{N}(x_{N}^{\st})=F^{\st},
\end{equation}
and $F^{\st}$ is calculated with the constraint $\sum_{i}x_{i}^{\st}=L$.
This solution is stable as long as $U''_{i}(x_{i}^{\st})>0$ for all $i$.

If all the units are identical, $U_{i}(x) = U(x)$,
the metastability regions of each module
(the range of forces for which the equation $U'_{i}(x)=F$ has several solutions) coincide.
Therefore, we obtain stationary branches corresponding
to $J$ unfolded units and $N-J$ folded units that have been analyzed
in detail in \cite{PCyB13,BCyP15}. If all the modules are not
identical, the metastability regions do not perfectly
overlap since the units are not equally strong: the weakest one is
that for which the equation $U'_{i}(x)=F$ ceases to have multiple
solutions for a smaller force value.

It is important to note that we can change all the forces
$U'_{i}(x_{i})$ to $V'_{i}(x_{i})=U'_{i} (x_{i})-F_{0}$ and $F$ to
$\varphi=F-F_{0}$, we have the same system \eqref{eq:7} but with
$V'_{i}$ and $\varphi$ instead of $U'_{i}$ and $F$, respectively.
Then, we may use the free energies for any common value of the force
$F_{0}$ and interpret the Lagrange multiplier as the excess force from
this value to be applied to the system\footnote{A similar result is
  also found if the length is controlled by using a device with a
  finite value of the stiffness $k_{p}$. A constant force only
  shifts the equilibrium point of a harmonic oscillator:
  $(q_{N}-L)$ must be substituted by $(q_{N}-L-F_{0}/k_{p})$.}.

\section{\label{sec:real_chain} The pulled chain}

Let us consider the pulling of our system.
We write the $i$-th-unit free energy as
\begin{equation}\label{eq:9}
 U_{i}(x) = U(x) +\xi\,\delta U_{i}(x),
\end{equation}
in which $U(x)$ is the ``main'' part, common to all the units, and
$\xi \delta U_{i}(x)$ represents the separation from this main
contribution. If all the units are perfectly identical,
$U_{i}(x)=U(x)$ for all $i$ or, equivalently, $\delta U_{i}(x)=0$.  In
principle, in an actual experiment, the splitting of the free
  energy in \eqref{eq:9} can be done if the free energy $U_{i}$ of
  each unit is known: we may define the common part as the ``average''
  free energy over all the units,
  $U(x)\equiv \overline{U}(x)\equiv N^{-1}\sum_{i=1}^{N} U_{i}(x)$,
  and $\xi \delta U_{i}(x)\equiv U_{i}(x)-\overline{U}(x)$. From a
  physical point of view, the dimensionless parameter $\xi>0$
  measures the importance of the heterogeneity in the free energies.
Our theory could be applied to a situation in which the free
  energy deviations $\delta U_{i}$ were stochastic and followed a
  certain probability distribution, for instance to represent the
  slight differences among very similar units, as done in
  \cite{BCyP15} to analyze the force-extension curves. In particular,
the forces $U'_{i}(x)$ in the evolution equations can also be split as
\begin{equation}
\label{eq:10}
 U'_{i}(x)=U'(x)+\xi \, \deltaf_{i}(x), \quad
\deltaf_{i}(x)\equiv\delta U'_{i}(x).
\end{equation}

\begin{figure}
  \centering
  \includegraphics[width=0.6 \textwidth]{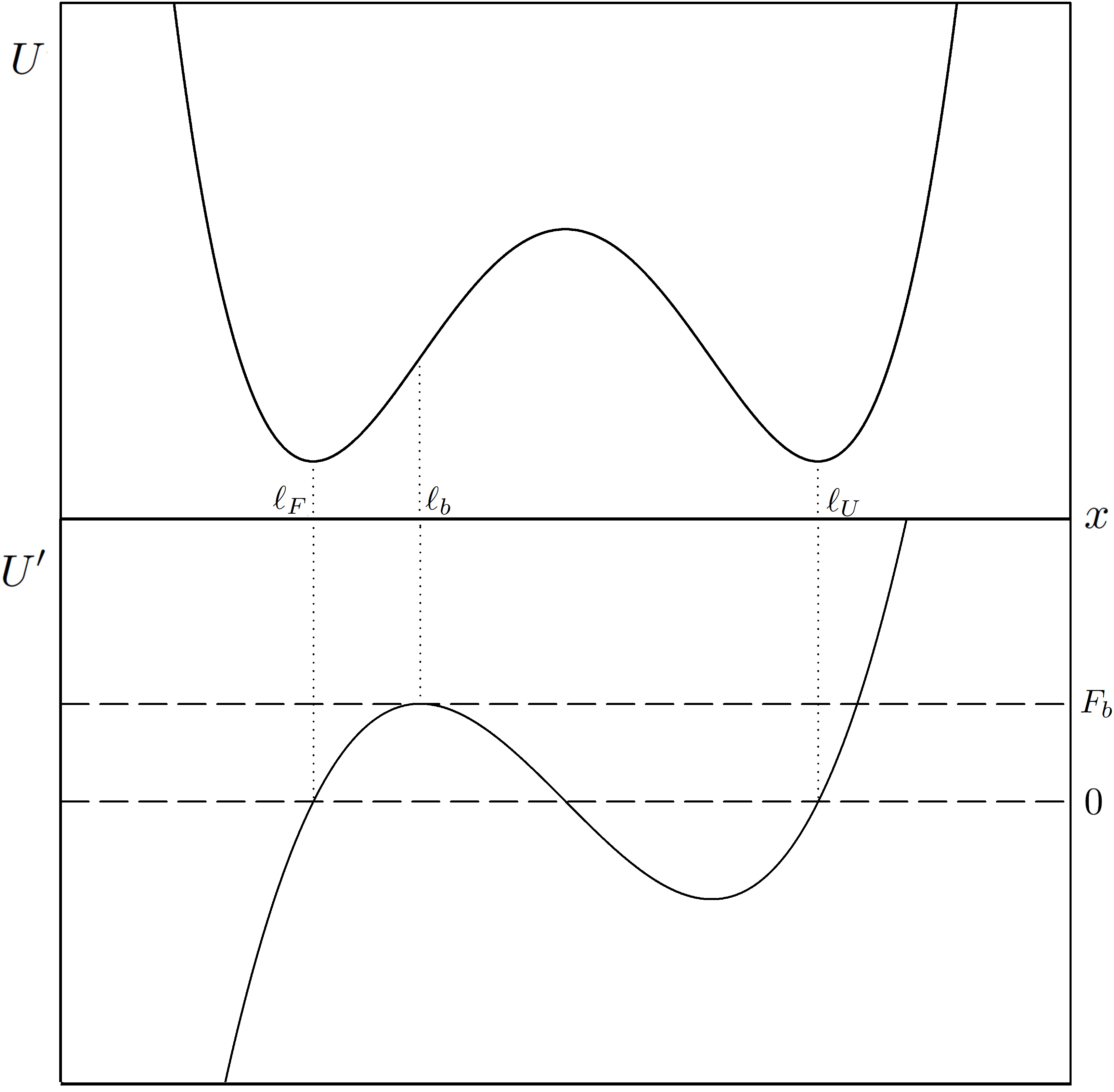}
  \caption{Qualitative behaviour of the main contribution to the free
    energy $U(x)$ at the critical force (top panel) and associated
    force $U'(x)$ (bottom panel) as a function of the extension.  The values of the
    lengths at the folded and unfolded minima are $\ell_{F}$ and
    $\ell_{U}$, respectively, whereas the threshold length $\ell_{b}$
    stands for the length corresponding to the limit of stability and
    $F_{b}$ is the corresponding force.}
  \label{fig:2}
\end{figure}

As already noted above, we can use the free energies for any
common value of the force $F_{0}$, and interpret $F$ as the extra
applied force from this value.  In what follows,
we consider $U(x)$ with two, equally deep, minima corresponding to the
folded (F)
and unfolded (U) configurations.  Figure \ref{fig:2} presents a
qualitative picture of the free energy and its derivative. The two
minima correspond to lengths $\ell_{F}$ and $\ell_{U}$, with
$\ell_{F}<\ell_{U}$. Also the point $\ell_{b}$
at which $U''(\ell_{b})=0$ is marked.

It is the condition $U''(\ell_{b}) = 0$ that essentially determines
the stability threshold, as it provides the limit force
$F_{b}=U'(\ell_{b})>0$ at which the folded basin ceases to exist for
the ``main'' potential.  
In the deterministic approximation considered here, thermal
  fluctuations are  neglected and, for $F<F_{b}$, the folded unit
  cannot jump over the free energy barrier hindering its unfolding: it
  has to wait until, at $F=F_{b}$,  the only possible extension is that
  of the unfolded basin. Of course, neglecting thermal noise restricts
  in some way the range of applicability of our results, see
  section~\ref{sec:critical_speed} for a more detailed discussion and
  also the numerical section \ref{sec:numerics}.

Keeping the above discussion in mind, now we analyze the limit of
stability of the different units. The asymmetry correction
  $\delta f_{i}$ shifts the threshold force for the different units.
The extension $x_{i,b}$ at which the $i$-th unit loses its stability
is obtained by solving the equation
$U''_i(x_{i,b}) = U''(x_{i,b}) + \xi \delta f'_{i}(x_{i,b}) = 0$,
which linearized in both the displacement $x_{i,b}-\ell_{b}$ and $\xi$
reads
\begin{equation}
U''(\ell_{b}) + U'''(\ell_{b})(x_{i,b} - \ell_{b}) + \xi \delta f'_{i}(\ell_b) = 0.
\end{equation}
Noting that $U''(\ell_{b}) = 0$, we get that
\begin{equation}
  \label{eq:10b}
  x_{i,b}=\ell_{b}-\xi \,\frac{\delta f'_{i}(\ell_{b})}{U'''(\ell_{b})}.
\end{equation}
See \ref{apa} for details. The corresponding force  is
\begin{equation}
  \label{eq:10c}
  F_{i,b}=F_{b}+\xi\,\deltaf_{i}(\ell_{b}),
\end{equation}
in which we have also dropped terms of the order of $\xi^{2}$.  Then,
units with $\deltaf_{i}(\ell_{b})<0$ ($\deltaf_{i}(\ell_{b})>0$) are
weaker (stronger) than average.

When the system is continuously pulled, the total length
of the system $L$ has been shown to be a good reaction
coordinate \cite{ACDNKMNLyR10}. Therefore, on physical grounds it
is reasonable to use $L$ to measure time and write the evolution equations
\eqref{eq:7} as
\begin{subequations}\label{eq:11}
  \begin{eqnarray}
\fl   \gamma v_{p} \frac{d x_{1}}{dL}  =  -U'(x_{1})+U'(x_{2})+\xi[-\delta
  f_{1}(x_{1})+\deltaf_{2}(x_{2})],
\\
\fl  \gamma v_{p} \frac{d x_{i}}{dL}  =  -2U'(x_{i})+U'(x_{i+1}) +U'(x_{i-1})
     +\xi[-2\deltaf_{i}(x_{i})+\deltaf_{i+1}(x_{i+1})+\delta
   f_{i-1}(x_{i-1})], \nonumber
\\
    \qquad\qquad\qquad\qquad\qquad\qquad\qquad\qquad\qquad\qquad\qquad 1<i<N,
\\
\fl  \gamma v_{p} \frac{d x_{N}}{dL}  =  -2U'(x_{N})+U'(x_{N-1})+F
+\xi[-2\delta
  f_{N}(x_{N})
  +\deltaf_{N-1}(x_{N-1})],
\\
\fl  F =\gamma v_{p}+U'(x_{N})+\xi\,\deltaf_{N}(x_{N}).
  \end{eqnarray}
\end{subequations}
Moreover, this change of variable makes the pulling speed $v_{p}$
  appear explicitly in the equations, allowing us to consider $v_{p}$
  as a perturbation parameter for slow enough pulling processes.

Now, we consider a system in which the asymmetry in the free energies
is small and which is slowly pulled. Thus, \eqref{eq:11}
is solved by means of a perturbative expansion in powers
of the pulling velocity $v_{p}$ and the disorder parameter $\xi$, that
is,
\begin{subequations}\label{eq:12}
  \begin{equation}
    \label{eq:12a}
    x_{i}(L)=x_{i}^{(0)}(L)+\xi \delta x_{i}(L)+v_{p} \Delta x_{i}(L),
  \end{equation}
  \begin{equation}
    \label{eq:12b}
    F(L)=F^{(0)}(L)+ \xi \delta F (L)+v_{p} \Delta F (L),
  \end{equation}
\end{subequations}
up to the linear order in both $v_{p}$ and $\xi$.

The zero-th (lowest) order corresponds to the chain of identical units
($\xi=0$) with a given constant length $L$ ($v_{p}=0$). Namely,
$x_{i}^{(0)}$ and $F^{(0)}$ obey the equations
\begin{subequations}\label{eq:13}
  \begin{eqnarray}
   0  =  -U'(x_{1}^{(0)})+U'(x_{2}^{(0)}),  \\
   0  =  -2U'(x_{i}^{(0)})+U'(x_{i+1}^{(0)})+U'(x_{i-1}^{(0)}), \quad 1<i<N, \\
   0  =  -2U'(x_{N}^{(0)})+U'(x_{N-1}^{(0)})+F^{(0)}, \\
   F^{(0)} = U'(x_{N}^{(0)}).
\end{eqnarray}
\end{subequations}
The solution of this system is straightforward,
\begin{equation}
  \label{eq:16}
  U'(x_{i}^{(0)})=F^{(0)},
\end{equation}
the force is equally distributed among all the units of the chain in
equilibrium, as expected.  If we start the pulling process from a
configuration in which all the units are folded and the force is
outside the metastability region (the usual situation), the units
extensions and the applied force are
\begin{equation}
  \label{eq:14}
  x_{i}^{(0)}=\ell\equiv \frac{L}{N},  \quad \forall i, \qquad F^{(0)}=U'(\ell),
\end{equation}
to the lowest order. To calculate the linear corrections in $\xi$ and
$v_{p}$, we have to substitute \eqref{eq:12} and \eqref{eq:14}
into \eqref{eq:11}, and equate terms
proportional to $\xi$ and $v_{p}$, respectively.
This is done below in
two separate sections: firstly, for the asymmetry contribution $\delta
x_{i}$ and, secondly, for the ``kinetic'' contribution $\Delta x_{i}$.

\subsection{\label{sec:asymmetry}Asymmetry correction}
All the modules are not characterized by the same free energy, and
here we calculate the first order correction introduced thereby.
The asymmetry corrections $\delta x_{i}$ obey the system of equations
\begin{subequations}\label{eq:18}
\begin{eqnarray}
\delta x_{2}-\delta x_{1}
=\frac{\deltaf_{1}(\ell)-\deltaf_{2}(\ell)}{U''(\ell)},\\
\delta x_{i+1}+\delta x_{i-1}-2\delta x_{i}  =
\frac{2\deltaf_{i}(\ell)-\deltaf_{i+1}(\ell)-\deltaf_{i-1}(\ell)}{U''(\ell)},
\quad  1<i<N \label{eq:18b}\\
\delta x_{N-1}-2 \delta x_{N}  =
\frac{2\deltaf_{N}(\ell)-\deltaf_{N-1}(\ell) -\delta F}{U''(\ell)}, \\
\delta F  =  U''(\ell) \delta x_{N}+\deltaf_{N}(\ell),
\end{eqnarray}
\end{subequations}
which is linear in the $\delta x_{i}$'s, and thus can be analytically
solved. It is clear that our expansion breaks down when
$U''(\ell)=0$. This was to be expected, since we know that the
stationary branch with all the modules folded is unstable when
$U''_{i}$ becomes negative for some unit $i$, and to the lowest order
this takes place when $U''(\ell)=0$.

The solution of the above system of difference equations is obtained
by standard methods \cite{ByO99},  with the
result
\begin{equation}
  \label{eq:19}
  \delta x_{i}=\frac{\overline{\deltaf}(\ell)-\deltaf_{i}(\ell)}{U''(\ell)}, \;
  \forall i,
  \quad \delta F=\overline{\deltaf}(\ell)=\frac{1}{N} \sum_{i=1}^{N} \delta f_i(\ell).
\end{equation}
See \ref{ap1} for more details. Interestingly, the force is homogeneous across the chain, since
  to first order in $\xi$ we have that
\begin{equation}
  U'_{i}(x_{i}) =U'(x_{i}^{(0)})+\xi [U''(x_{i}^{(0)})\delta x_{i}+
  \deltaf_{i}(x_{i}^{(0)})]  =  U'(\ell)+\xi  \overline{\deltaf}(\ell)=F^{(0)}+\xi \delta F.
\label{eq:20}
\end{equation}
This is nothing but the stationary solution \eqref{eq:17}, up
  to first order in the disorder\footnote{If the zero-th order free
    energy were the average of the $U_{i}$'s, no correction for the
    Lagrange multiplier (applied force) would appear to the first
    order. This is logical, up to the first order the force expression
    coincides with the spatial derivative of the average potential,
    that is,
    $F^{(0)}+\xi \delta F=U'(\ell)+\xi
    \overline{\deltaf}(\ell)=\overline{U'}(\ell)$.}.
  Moreover, (\ref{eq:19}) implies that there are units with
  $\delta x_{i}>0$ and others with $\delta x_{i}<0$, depending on the
  sign of $\overline{\deltaf}(\ell)-\deltaf_{i}(\ell)$. This is a
  consequence of our perturbation expansion, since
  $\sum_{i}x_{i}^{(0)}=L$ for all times, as given by
  (\ref{eq:14}), and thus $\sum_{i}\delta x_{i}=0$.

Let us remember that we denote by $\ell_{b}$ the value of the
extension at which the common main free energy reaches its limit of
stability, see figure \ref{fig:2}. Taking into account only the
asymmetry correction, it is the weakest unit that unfolds first, since
the most negative $\deltaf_{i}(\ell)$ leads to the largest positive
$\delta x_{i}$ and then it is the one that first verifies the
condition $x_{i}=\ell+\xi\delta x_{i}=\ell_{b}$ (for a more detailed
discussion, see \ref{apa}). An alternative way of looking at this
is to recall that the force corresponding to the limit of stability
is smallest for the weakest unit: since the force is homogeneously
distributed along the chain, it is the weakest module that first
reaches its stability threshold.

\subsection{Correction due to the finite pulling speed
\label{sec:pulling_speed}}
Now we look into the ``kinetic'' correction due to the finite pulling
speed $v_{p}$. The zero-th order solution is given by
\eqref{eq:14}, so that $dx_{i}^{(0)}/dL=N^{-1}$ for all $i$, and
we have
\begin{subequations}
  \label{eq:21}
\begin{eqnarray}
\Delta x_{2}-\Delta x_{1}=\frac{\gamma}{N U''(\ell)},\\
\Delta x_{i+1}+\Delta x_{i-1}-2\Delta x_{i}  =
\frac{\gamma}{N U''(\ell)}, \quad 1<i<N, \label{eq:21b}\\
\Delta x_{N-1}-2 \Delta x_{N}  =
\frac{1}{U''(\ell)}\left[\frac{\gamma}{N}-\Delta F \right], \\
\Delta F  =  \gamma + U''(\ell)\Delta x_{N}.
\end{eqnarray}
\end{subequations}
The solution to this system of linear difference equations \cite{ByO99} is
\begin{subequations}\label{eq:22}
\begin{eqnarray}
  \label{eq:22a}
\Delta x_{i}=\frac{\gamma}{2NU''(\ell)} \left[i(i-1)-\frac{(N+1)(N-1)}{3}
\right],\\
\label{eq:22b}
\Delta F=\frac{(N+1)(2N+1)\gamma}{6N}.
\end{eqnarray}
\end{subequations}
See \ref{ap1} for more details. Again, $\sum_{i}\Delta x_{i}=0$ because the zero-th order solution
  (\ref{eq:14}) gives the total length, $\sum_{i}x_{i}^{(0)}=L$ for
  all times.
\eqref{eq:22a} is reasonable on intuitive
grounds: the kinetic correction $\Delta x_{i}$ increases with $i$
because the last module is the one that is actually pulled. Therefore,
on the basis of only the kinetic correction, it is the last module
that would unfold first because $\Delta x_{N}$ is the largest.  Thus,
the condition $x_{i}=\ell+v_{p}\Delta x_{i}=\ell_{b}$ is first
verified for $i=N$.

It is interesting to note that the force was equally distributed for
the asymmetry correction, as expressed by \eqref{eq:20}, but this
is no longer true if we incorporate the kinetic correction.  Up to the
the first order, $U'_{i}(x_{i})=U'(x_{i}^{(0)}+\xi\delta x_{i}+v_{p}\Delta
x_{i})+\xi \deltaf_{i}(x_{i})\simeq U'(\ell)+\xi
\overline{\deltaf}(\ell)+v_{p}U''(\ell) \Delta x_{i}$. Therefore, the force
$U'_{i}(x_{i})$ depends on the unit $i$: for all times, it is smaller
the further from the pulled unit we are. Again, there is an
alternative way of understanding why the last unit would unfold first
if we were considering perfectly identical units ($\xi=0$): for any
time, it would be the last unit that suffered the largest force and
thus the first that reached their common limit of stability $F_{b}$.

\subsection{\label{sec:critical_speed}The critical velocities}
If the last unit is not the weakest, there is a competition between
the asymmetry and the kinetic corrections. For very low pulling
speeds, in the sense that $v_{p}/\xi \to 0$, the term
  proportional to $v_{p}$ can be neglected and it is the weakest unit
  (the one with the largest $\delta x_{i}$) that unfolds first, as
  discussed in section \ref{sec:asymmetry}. On the other hand, for
  very small disorder, in the sense that $\xi/v_{p} \to 0$, the
  term proportional to $\xi$ is the one to be neglected and it is the
  last unit (the one with the largest $\Delta x_{i}$) that unfolds
  first, as also discussed in
  section \ref{sec:pulling_speed}. Therefore, different unfolding
    pathways are expected as the pulling speed changes.

Collecting all the contributions to the extensions, we have that
\begin{equation}
x_{i}=\ell+ \frac{\xi\overline{\deltaf}(\ell)-v_{p}\gamma\dfrac{N^2-1}{6N}}{U''(\ell)}
   + \frac{v_{p}\gamma \dfrac{ i(i-1)}{2N}-\xi \deltaf_{i}(\ell)}{U''(\ell)} .
\label{eq:23}
\end{equation}
We have rearranged the terms in $x_{i}$ in such a way that the first
two terms on the rhs are independent of the unit $i$, all the
dependence of the length of the module on its position across the
chain has been included in the last term. Note that we are expanding
the solution in powers of $v_{p}$ around the ``static'' solution,
  which is obtained by putting $v_{p}=0$ in \eqref{eq:23}. Thus, the
  ``static'' solution corresponds to the stationary one the
  system would reach if we kept the total length constant and equal to
  its instantaneous value at the considered time.  It is essential to
  realise that \eqref{eq:23} is only valid for very slow pulling, as
  long as the corrections to the ``static'' solution are small, and
this is the reason why the limit of stability is basically unchanged
as compared to the static case.  In order to be more precise, we refer
to this kind of very slow pulling as \textit{adiabatic} pulling. A main
result of our paper is that, even for the case of adiabatic pulling,
there appear different unfolding pathways depending on the value of
the pulling speed.

In the adiabatic limit we are considering here, the pulling
  process has to be slow enough to make the system move very close to
  the stationary force-length branches, but not so slow that the
  system has enough time to escape from the folded basin. As discussed
  in \cite{BCyP15}, there is an interplay between the pulling velocity
  and thermal fluctuations. For very slow pulling velocities, the
  system has enough time to surpass the energy barrier separating the
  two minima, which leads to the typical logarithmic dependence of the
  ``unfolding force'' $F_{U}$ on the pulling speed, specifically
  $F_{U}\propto (\ln v_{p})^{a}$ \cite{RGPCyS13,DHyS06}.\footnote{The
    parameter $a$ is of the order of unity, its particular value
    depends on the specific shape of the potential (linear-cubic,
    cuspid-like, ...)  considered \cite{DHyS06}.  }  On the other
  hand, as already argued at the beginning of section
  \ref{sec:real_chain}, for \textit{adiabatic pulling}, the units
  unfold not because they are able to surpass the free energy barrier
  but because the folded state ceases to exist at the force $F_{b}$
  corresponding to the upper limit of the metastability region.

The unit that unfolds first is the one for
which $x_{i}=\ell_{b}$ for the shortest time. In light of the above,
it is natural to investigate whether it is possible to determine
which module is the first to unfold for a given pulling speed. To
put it another way, we would like to calculate the ``critical''
velocities which separate the velocity intervals in which a specific
module unfolds first. Let us assume that, for a given pulling speed
$v_{p}$, it is the $i$-th module that unfolds first. All the modules
$j$ to its left, that is, with $j<i$, will not open first if the
pulling velocity is further increased because the difference between
the kinetic corrections $\Delta x_{i}-\Delta x_{j}$ increases with
$v_{p}$. Therefore, the first module $j$ to unfold when the velocity
is sufficiently increased it is always to its right.  The velocity
$v^{i}(j)$ for which each couple of modules $(i,j)$, $j>i$, reach
  simultaneously the stability threshold is determined by the
  condition
\begin{equation}
  \label{eq:stab}
  x_{i}(\ell_{c})=x_{j}(\ell_{c}) = \ell_{b}
\end{equation}
\eqref{eq:stab} determines both the value of $\ell_{c}$
(or time $t_{c}$) at which the stability threshold is reached and the
relationship between $v_{p}$ and $\xi$.  \eqref{eq:23} implies
that
\begin{equation}
  \label{eq:25a}
  -\xi \delta f_{i}(\ell_{c})+\gamma v^{i}(j)
  \frac{i(i-i)}{2N} =-\xi \deltaf_{j}(\ell_{c})+
  \gamma v^{i}(j) \frac{j(j-1)}{2N}.
\end{equation}
We already know that the length corresponding to the limit of
stability is very close to the threshold length $\ell_{b}$, its
distance thereto being of the order of $\sqrt{\xi}$, as shown in
 \ref{apa}. Therefore, to the lowest order, $\ell_{c}$ can be
approximated by $\ell_{b}$, and we get
\begin{equation}
  \label{eq:vcij}
  \frac{\gamma v^{i}(j)}{\xi}=\frac{2N[\deltaf_{j}(\ell_{b})-
    \deltaf_{i}(\ell_{b})]}{j(j-1)-i(i-1)}, \quad j>i.
\end{equation}
Clearly, the minimum of these velocities is the one that matters:
  Let us denote by $j^{(i)}_{\text{min}}$ the position of the module for which
  $v^{i}(j)$ reaches its minimum value $v_{\text{min}}^{i}$,
\begin{equation}
  \label{eq:vci}
  v_{\text{min}}^{i}=v^{i}(j^{(i)}_{\text{min}})=\min_{j} v^{i}(j)
\end{equation}
for $v_{p}$ just below $v_{\text{min}}^{i}$, it is the $i$-th
  module that unfolds first, but for $v_{p}$ just above
  $v_{\text{min}}^{i}$, it is the $j$-th module that unfolds first.
  Let us denote the weakest module by $\alpha_{1}$, that is,
  $\deltaf_{i}(\ell_{b})$ is smallest for $i=\alpha_{1}$. If $v_p$ is
  smaller than $v_{\text{min}}^{\alpha_{1}}$, the first unit to reach the
  stability limit is the weakest one. Then, we rename the latter
  velocity $v_c^{(1)}$, that is,
\begin{equation}
v_c^{(1)} = v_{\text{min}}^{\alpha_{1}} , \quad
\deltaf_{\alpha_{1}}(\ell_{b})=\min_{i}\deltaf_{i}(\ell_{b}),
\label{eq:vcw}
\end{equation}
because it is the first one of a (possible) series of critical
  velocities separating different unfolding pathways, see below.

Let us denote by $\alpha_{2}$ the module which unfolds first in the
  ``second'' velocity region, $v_{p}$ just above $v_{c}^{(1)}$, that
  is, $\alpha_{2}=j_{\text{min}}^{(\alpha_{1})}$. This unit ceases to be the
first to unfold for the velocity
\begin{equation}
v_c^{(2)} = v_{\text{min}}^{\alpha_{2}}
\label{eq:vc1}
\end{equation}
The successive changes on the unfolding pathway take place at the critical
velocities
\begin{equation}
v_c^{(k)} = v_{\min}^{\alpha_{k}},
\label{eq:vck}
\end{equation}
in which $\alpha_{k+1}=j_{\text{min}}^{(\alpha_{k})}$.
This succession ends when $\alpha_{k+1}=N$: in that case, for
$v_{p}>v^{(k)}_c$, the  first unit to unfold is
always the pulled one.
This upper critical velocity $v_c^{\text{end}}$
can be computed in a more direct way,
\begin{equation}
v_c^{\text{end}} = \max_{j} v_{\text{min}}^{j}(N).
\label{eq:vce}
\end{equation}
Consistency of the theory requires that $v_c^{(k+1)}>v_c^{(k)}$; a
short proof is presented in \ref{apb}.

We have a trivial case for $\alpha_{1}=N$, when the pulled unit is
precisely the weakest and it is always the first to unfold for any
pulling speed. The simplest nontrivial case appears when all the
modules has the same free-energy with the exception of the weakest,
and $\alpha_{1} \neq N$, \eqref{eq:vcij},
\eqref{eq:vcw} and \eqref{eq:vce} reduce to
\begin{equation}
  \label{eq:26a}
  \frac{\gamma v^{(1)}_{c}}{\xi} =
  \frac{\gamma v^{\text{end}}_{c}}{\xi}=\frac{2N[\deltaf_{N}(\ell_{b})- \deltaf_{\alpha_{1}}(\ell_{b})]}{N(N-1)-\alpha_{1}(\alpha_{1}-1)}.
\end{equation}
Note that the situation is quite simple, since there exist a single
critical velocity $v_c = v_c^{(1)} = v_c^{\text{end}}$.  For
$v_p < v_c$ the weakest module unfolds first whereas for $v_p > v_c$
the last one unfolds first.  In general, when the units have different
free energies the situation may be more complex, as shown in the
previous paragraph. There appear intermediate critical velocities,
which define pulling speed windows where neither the weakest
unit nor the last one is the first to unfold.  In order to obtain
these regions, we need to recursively evaluate \eqref{eq:vck}.

\section{\label{sec:numerics}Numerical results}

Throughout this numerical section, we check the agreement between our
theory and the numerical integration of the evolution
equations. Firstly, we discuss the validity of the simplifications
introduced in the development of the theory, namely (i) negligible
thermal noise and (ii) perfect length-control. Secondly, we look into
the critical pulling speed, showing that there appears such a critical
speed in the simulations and comparing this numerical value to the
theory developed before.

We consider a system composed of $N=4$ unfoldons,
such as the maltose binding protein \cite{GMTCyC14},
each one characterized by a quartic bistable free energy.
In reduced variables, the free energies have the
form $U_{i} (x)= \epsilon_{i} U(x)$, where
\begin{equation}
\label{eq:30}
  U(x)= \frac{1}{4} \left[(x-\sigma)^{2}-a^{2}\right]^{2},
\end{equation}
with $\epsilon_{i}=1$ for $i\neq 1$, $\epsilon_{1}< 1$,
$\sigma=0$ and $a=3$\footnote{Here, the value of $\sigma$ is
  different from the one in \cite{GMTCyC14} ($\sigma=8$). Its
  only effect is a shift of the origin of the extensions, our choice
  implies that a positive (negative) sign of the extension corresponds
  to an unfolded (folded) configuration.}.
The value of the friction coefficient is, also in reduced variables,
$\gamma=1$. We use these dimensionless reduced variables in order to
make it easier to compare our results to those in \cite{GMTCyC14}.
The function \eqref{eq:30} is one of the simplest, but
reasonable, choice to describe the free energy of different unfoldons
of the same protein domain.
Using our notation, we have
\begin{equation}
\deltaf_{i}(x)=0, \; i\neq 1, \qquad \deltaf_{1}(x)=-\xi U'(x),
\end{equation}
with $\xi=1-\epsilon_1$.
\eqref{eq:10b} and \eqref{eq:10c} give us the limits of
stability up to first order in the asymmetry $\xi$,
\begin{equation}
  \label{eq:34}
  x_{i,b}=\ell_{b},\;\forall i, \quad F_{i,b}=F_{b}, \; i\neq 1, \qquad F_{1,b}=(1-\xi_{1})F_{b}.
\end{equation}
For this simple example, \eqref{eq:34} is exact.
The weakest unit is the first one, because $F_{1,b}$ is the minimum
value of the force at the limit of stability. For the values of
  the parameters we are using, $\ell_{b}=a/ \sqrt{3}=1.73$ and
$F_{b}= U'(\ell_b) = 2\sqrt{3} a^3/9 = 10.4$. Since we are writing the
free energies for a common given value of the force, we are assuming
that all the units have their two minima equally deep at the same
force. This assumption is made to keep things simple: the main
ingredient for having an unfolding pathway that depends on the pulling
speed is to have different values of the forces $F_{i,b}$ at the
stability threshold for the different units.

In the case we are considering, the weakest unit is the first one,
while the others share the same free energy. This means that we have
the simplest scenario for the critical velocity in our theoretical
approach: it is always the weakest (for $v_{p}<v_{c}$) or the last
(for $v_{p}>v_{c}$) unit that opens first, as discussed at the end of
the previous Section. Here, (\ref{eq:26a}) for $\alpha_{1}=1$ and
$N=4$ reduces to
\begin{equation}\label{eq:simple}
\frac{\gamma v_{c}}{\xi}=\frac{2}{3} F_{b}.
\end{equation}

\begin{figure}
\centering
\includegraphics[width=0.7\textwidth]{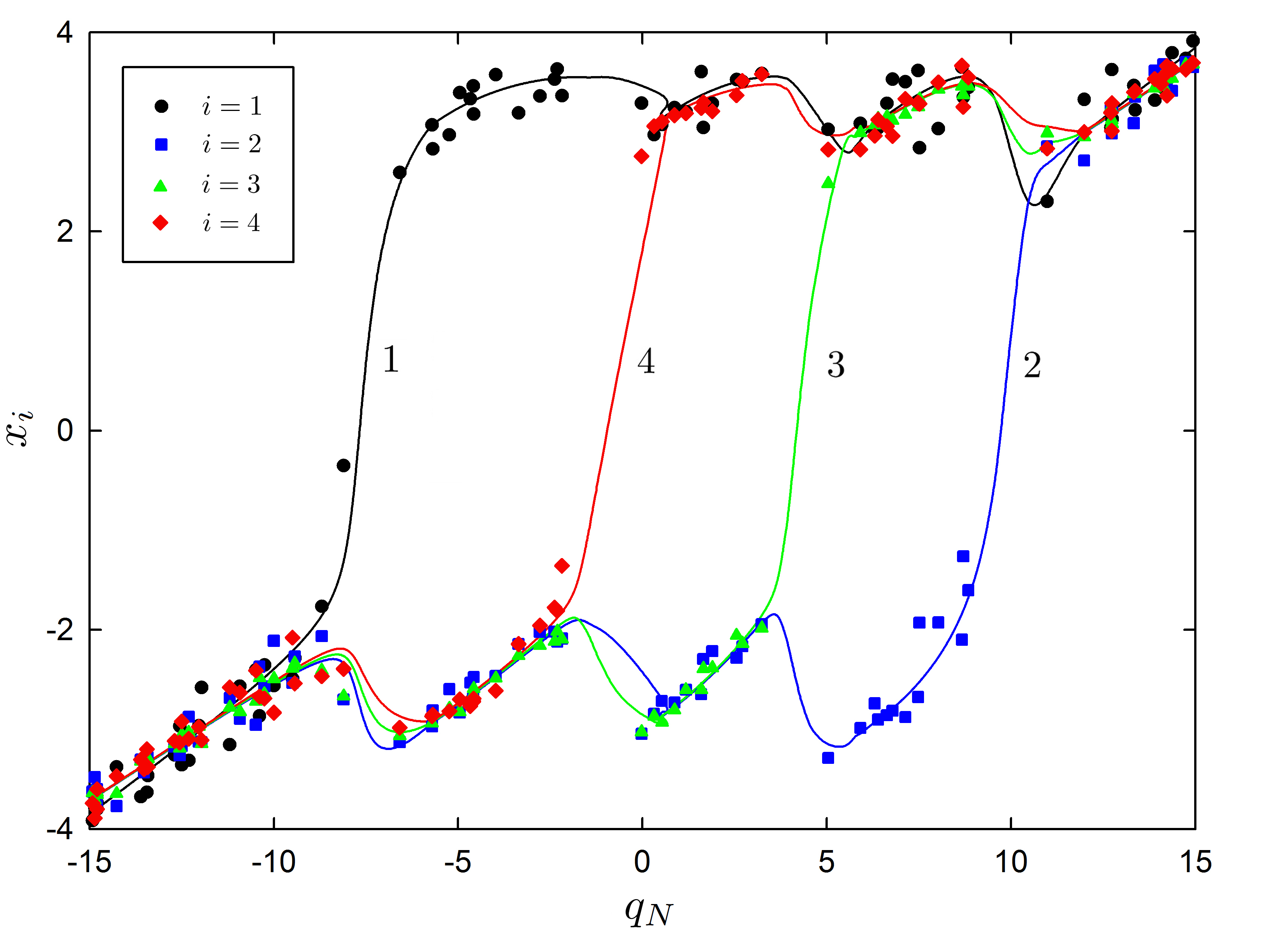}
\caption{\label{fig:3} Evolution of the extensions of the different
  units in a pulling experiment as a function of the length of the
  system $q_{N}$. The pulling speed is $v_{p}=0.38$ and the
    length-control device has a stiffness $k_{p}=5$.
  The symbols correspond to a typical realization of the
    Langevin process \eqref{eq:5} with $T=1$, whereas the lines
  correspond to the deterministic (zero noise) approximation.  }
\end{figure}

To start with, we consider the relevance of the noise terms in
\eqref{eq:5}. In figure \ref{fig:3}, we plot the integration of the
Langevin equations together with the deterministic approximation
\cite{vK92} for a concrete case: the free energy of the first
  unit corresponds to $\epsilon_{1}=0.8$ ($\xi=0.2$), the
  stiffness of the device controlling the length is $k_{p}=5$, the
  temperature is $T=1$ and the pulling speed is $v_{p}=0.38$. For
these values of the parameters, taken from \cite{GMTCyC14}, the
critical velocity in \eqref{eq:simple} is $v_{c}=1.4$, so we are
considering a subcritical velocity, $v_{p}<v_{c}$. Thermal
fluctuations are small, and thus the same unfolding pathway is
observed in the deterministic and the majority of the stochastic
  trajectories.

\begin{figure}
  \centering
\includegraphics[width=0.49\textwidth]{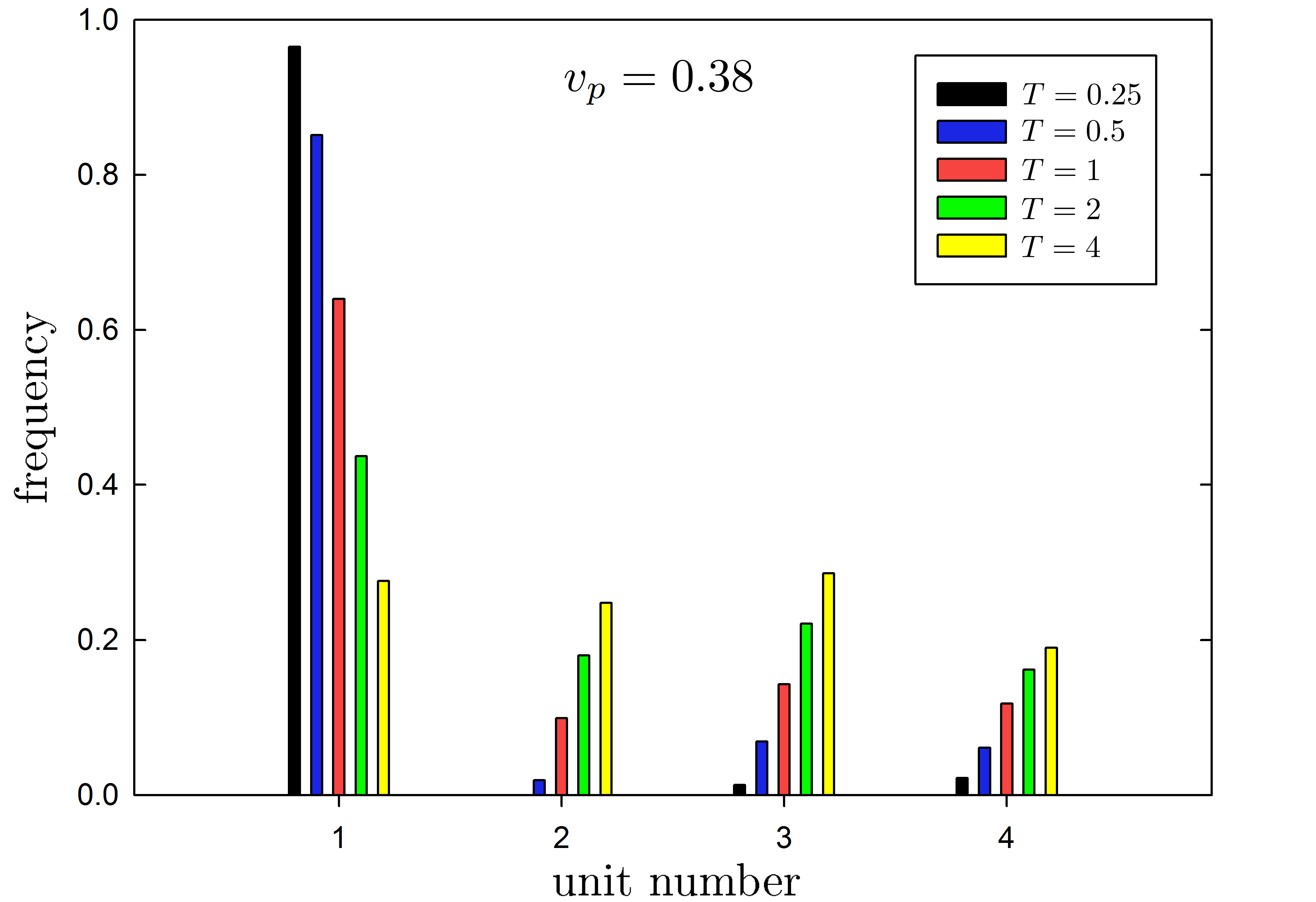}
\includegraphics[width=0.49\textwidth]{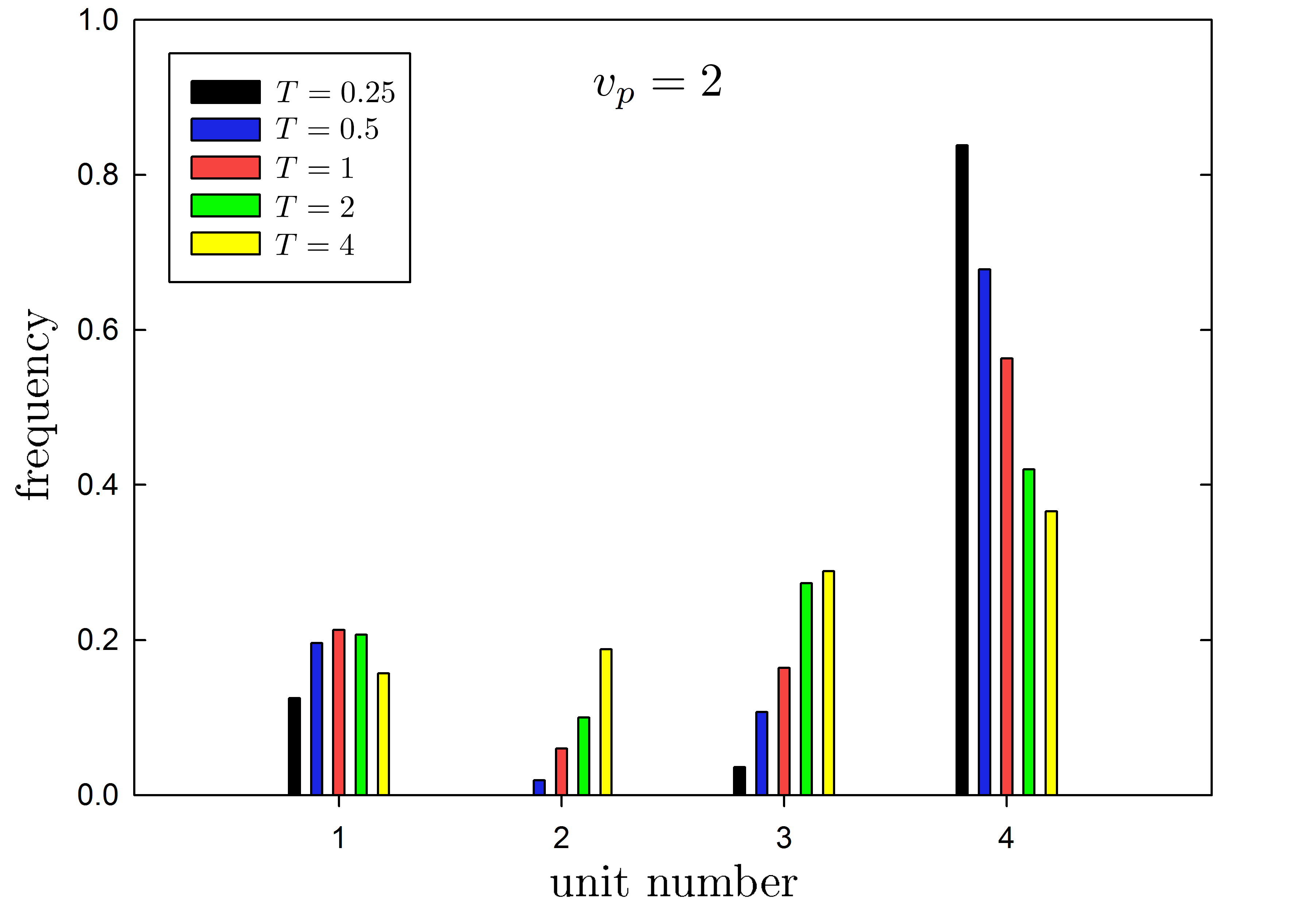}
\caption{Frequency with which each of the units unfolds first when the
  Langevin equations with perfect length-control are integrated for
  different values of the temperature.  (Left) Numerical frequencies
  obtained in $1000$ trajectories, for a subcritical pulling speed
  $v_{p}=0.38<v_{c}$.  (Right) The same as in the left panel, but for
  a supercritical pulling speed $v_{p}=2>v_{c}$. As the temperature
  decreases, the frequency of the deterministic unfolding pathway
  approaches unity in both cases.}
  \label{fig:new}
\end{figure}

Let us consider in more detail the relevance of thermal noise:
  from a physical point of view, it may be inferred by looking at the
  height of the free energy barrier at the critical force in terms of
  the thermal energy $k_{B}T$. For the values of the parameters we are
  using, this barrier is around $20$ in reduced units.  This explains
  why thermal noise is basically negligible in figure \ref{fig:3}, in
  which $T=1$. If the temperature is decreased to $T=0.25$, the
  barrier is so high, around $80$ times the thermal energy, that
  essentially all the stochastic trajectories coincide with the
  deterministic one. On the other hand, if the temperature is
  increased to $T=4$, the barrier in only a few, around $5$, times the
  thermal energy, and we expect that the deterministic approximation
  ceases to be valid. In order to further clarify this point, we
  present figure~\ref{fig:new}. Both panels display bar graphs with
  the frequencies with which each unit unfolds first in the stochastic
  trajectories obtained over $1000$ trajectories of the Langevin
  equations \eqref{eq:5} with perfect length control and different
  values of the temperature. In the left panel, a subcritical velocity
  $v_{p}=0.38<v_{c}$ is considered, so that the weakest (first) unit
  is expected to unfold first. In the right panel, the numerical data
  for a supercritical velocity $v_{p}=2>v_{c}$ are shown, for which the
  pulled (fourth) unit would unfold first. The effect of thermal noise
  is quite similar in both cases. For the low temperature $T=0.25$,
  the frequency of the deterministic pathway is close to unity and,
  for the temperature in figure~\ref{fig:3}, $T=1$, its frequency is
  still very large, clearly larger than any of the others. On the
  other hand, for the higher value of the temperature, $T=4$, thermal
  noise can no longer be neglected.

\begin{figure}
  \centering
  \includegraphics[width=0.49  \textwidth]{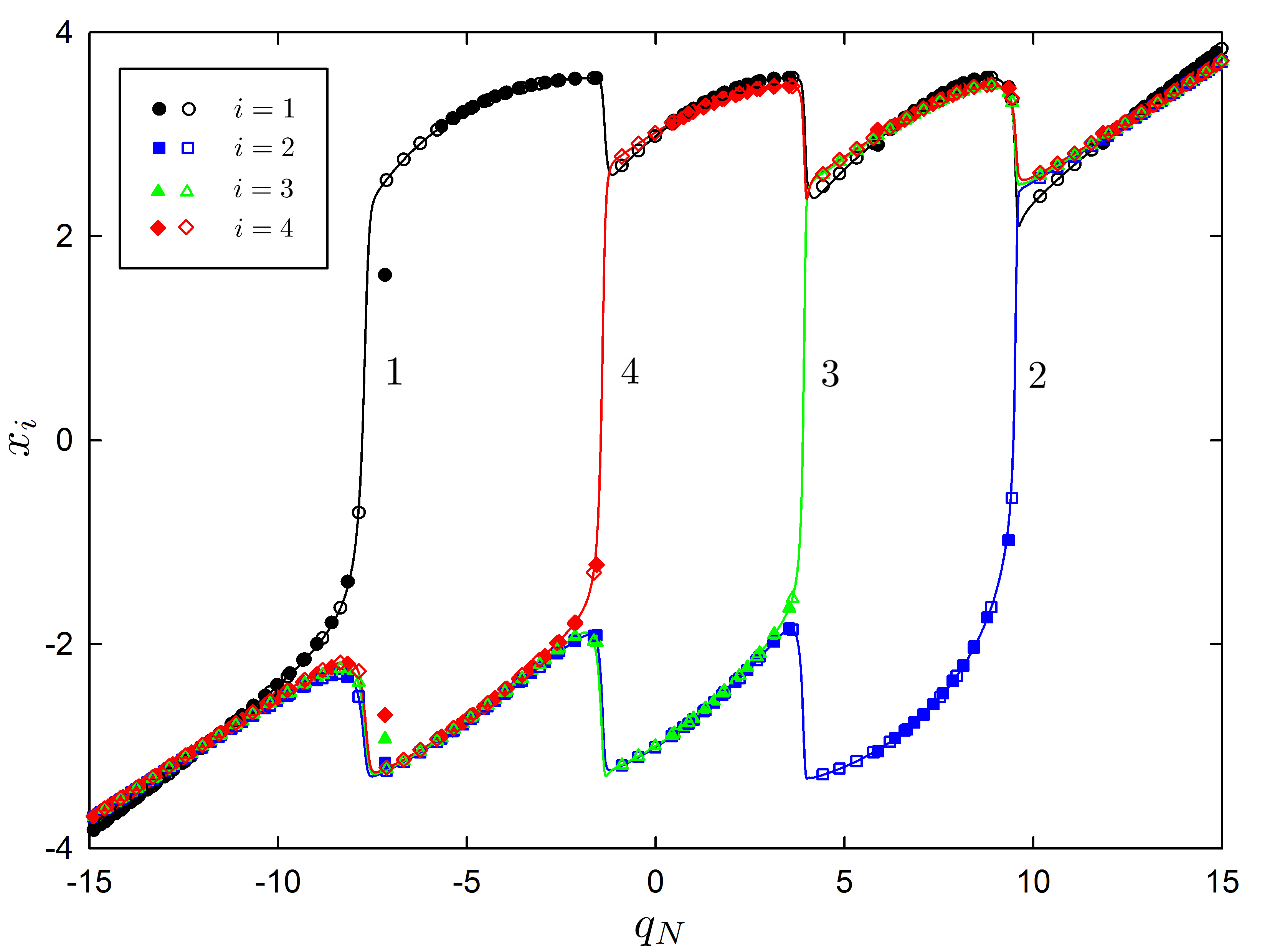}
  \includegraphics[width=0.49 \textwidth]{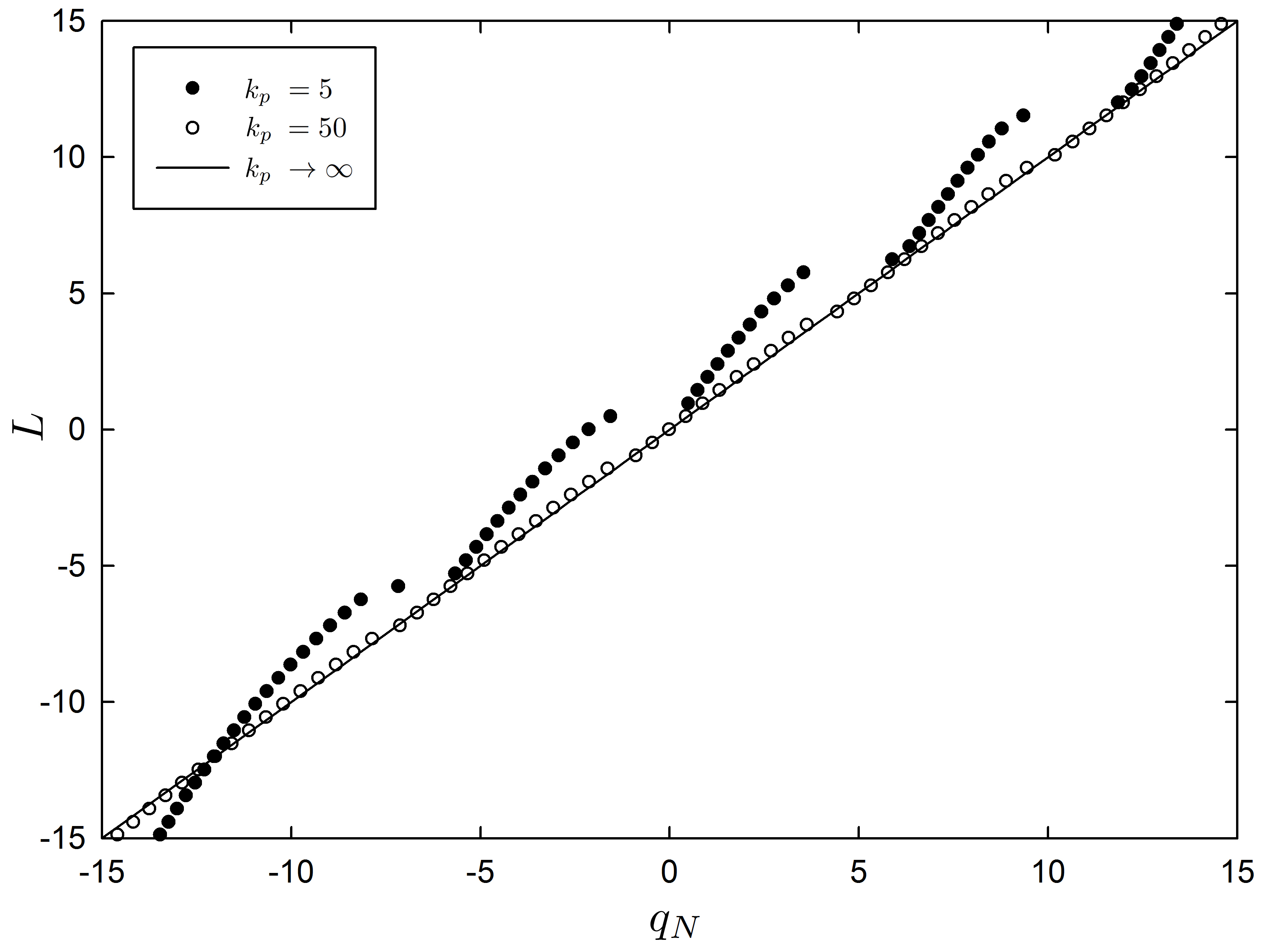}
  \caption{(Left) Evolution of the extensions of the different units
    in a pulling experiment as a function of the length of the system
    $q_{N}$. The symbols correspond to the integration of the
    deterministic equations, for $k_{p}=5$ (filled symbols) and
    $k_{p}=50$ (empty symbols), whereas the line correspond to the
    limit $k_{p}\to\infty$. The pulling speed is the same as in
      figure~\ref{fig:3}, that is, $v_{p}=0.38$. (Right) Comparison
    between the desired and actual lengths, $L$ and $q_{N}$, for the
    different values of the stiffness considered in the top panel. It
    is observed that the length control improves as $k_{p}$ increases.
  }
  \label{fig:4}
\end{figure}

In the following, we restrict the analysis to the physically
  relevant case in which the deterministic approximation gives a good
  description of the first unfolding event. In figure \ref{fig:4}
(left panel), we look into the same pulling experiment as before, but
now we compare the deterministic evolution of the extensions for two
finite values of the stiffness to the $k_{p}\to\infty$
limit. Consistently with our expectations, the unfolding pathway is
not affected by this simplification. Nevertheless, the control of the
length of course improves as $k_{p}$ increases (see right panel).
Although for the smaller values of $k_{p}$ the length is not perfectly
controlled, the curves in the left panel, which correspond to
different values of $k_{p}$, are almost perfectly superimposed when
plotted as a function of the real length of the system $q_{N}$ (but
not of the desired length $L$). This means that the real length
$q_{N}$ is a good reaction coordinate, as already said in section
\ref{sec:real_chain}.

\begin{figure}
  \centering
  \includegraphics[width=0.7  \textwidth]{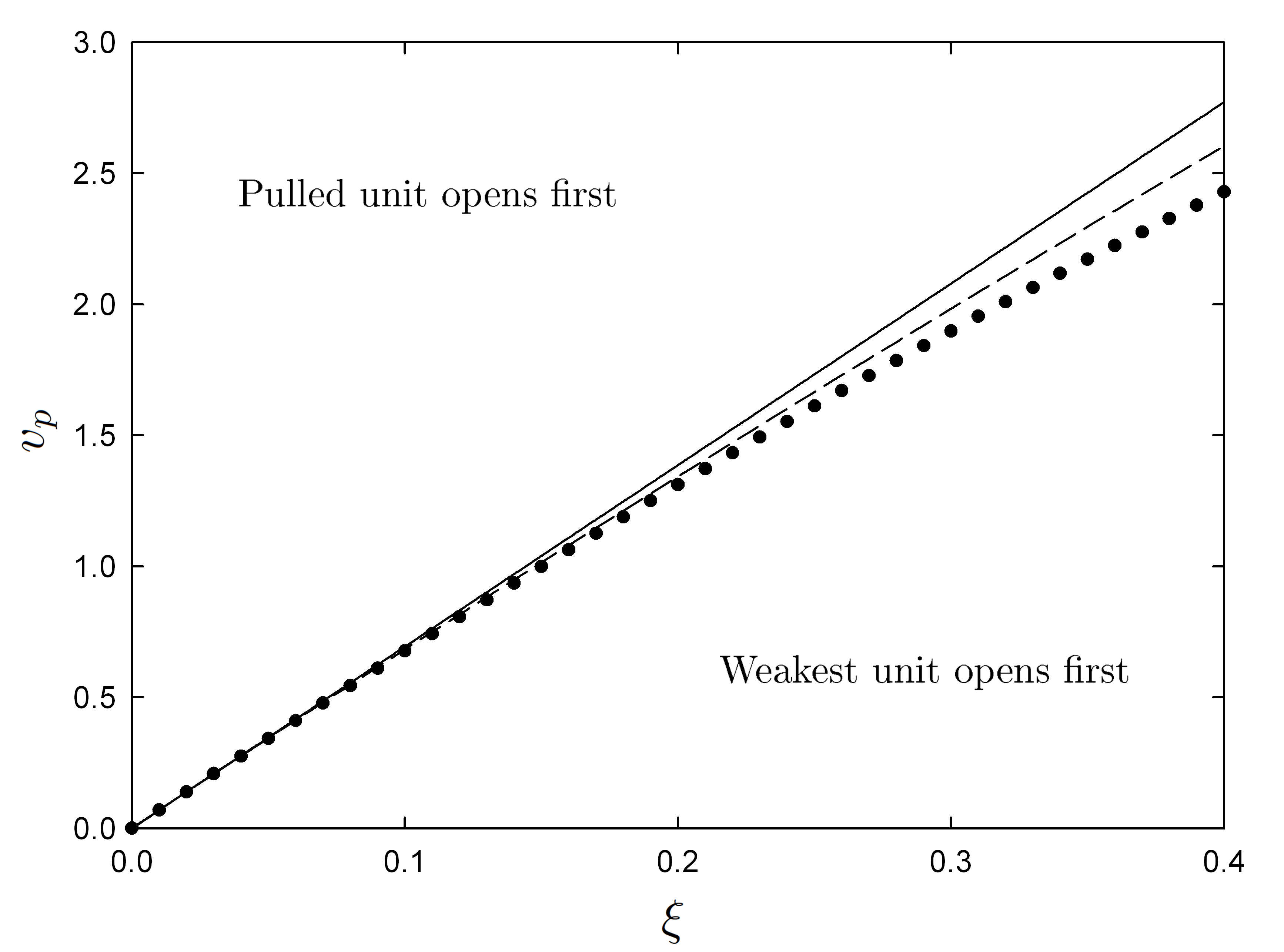}
  \caption{Phase diagram for the unfolding pathway in the pulling
    velocity-asymmetry plane. Two well-defined zones are separated by
    the curve giving the critical velocity $v_{c}$ as a function of
    the asymmetry $\xi$ in the free energy of the first unit. The
    numerical values for $v_{c}$ (circles) are compared to the
    theoretical expression (solid line), (\ref{eq:simple}). The
    dashed line corresponds to the alternative approach discussed in
    the text, which improves the agreement between the numerical
    results and the theory for $\xi>0.1$. Error bars have been omitted
    because they are smaller than point size.  }
  \label{fig:5}
\end{figure}

We have integrated the deterministic approximation  \eqref{eq:7} (zero noise) of the
Langevin equations  for different values of the
pulling speed, and extracted from them the numerical value of the
critical velocity as a function of the asymmetry
$\xi=1-\epsilon_{1}$. In order to obtain this numerical prediction, we initially set $v_{p}$ equal to the theoretical critical velocity given by \eqref{eq:simple}. Then, we recursively shift it by a small amount $\delta v_{p}$, such that $\delta v_{p}/v_c=0.0001$, until the pathway changes. We compare the values so obtained to the
theoretical expression \eqref{eq:simple}, in figure \ref{fig:5}. We
find an excellent agreement for $\xi\lesssim 0.1$, for $\xi>0.1$ there
appears some quantitative discrepancies.  These discrepancies stem
from two points: (i) the perturbative expansion used for
obtaining \eqref{eq:26a} from \eqref{eq:stab} and (ii) the
intrinsically approximate character of \eqref{eq:stab}, since
$\ell_{b}$ gives rigorously the limit of stability only for the static
case $v_{p}=0$.  Therefore, we have looked for the solution of
\eqref{eq:stab} in the numerical integration of the deterministic
equations. This is the dashed line in figure \ref{fig:5}, which
substantially improves the agreement between theory and numerics
because we have eliminated the deviations arising from point (i)
above. In fact, for the case we have studied in the previous
figures, which corresponds to a not so small asymmetry $\xi=0.2$, the
improved theory gives an almost perfect prediction for the critical
velocity.

Now we consider a more realistic potential for the units, which has
been introduced by Berkovich et al.~for modelling the unfolding of
single-unit I27 and ubiquitin proteins in AFM experiments
\cite{ber10,ber12}. Moreover, it has also been used to investigate the
stepwise unfolding of polyproteins in force-clamp conditions
\cite{BCyP14} and their force-extension curves in
  \cite{BCyP15}. At zero force, it reads
\begin{equation}
U(x)  =  U_0 \left[\left(1-e^{-2b(x-R_c)/R_c}\right)^2-1\right]
+\frac{k_BTL_c}{4P}\left(\frac{1}{1-\frac{x}{L_c}}-1-\frac{x}{L_c}
+\frac{2x^2}{L_c^2}\right),  \label{eq:berkovich}
\end{equation}
that is, it is the sum of a Morse and a worm-like-chain potentials,
representing the enthalpic and the entropic contributions to the free
energy, respectively \cite{ber10,ber12}. We take the values of the
parameters from \cite{BCyP15,ber10}, $P=0.4$nm (persistence length),
$L_c=30$nm (contour length), $T=300$K, $U_0=100$pN
nm($\sim\!\!  24 k_B T$), $R_c=4$nm, $b=2$. We measure force and
extensions in the units $[F]=100$ pN, $L_c=30$ nm,
respectively. Accordingly, dimensionless variables are introduced with
the definitions $\mu=U_0/(L_c[F])$, $\beta=2bL_c/R_c$, $\rho=R_c/L_c$,
$A=k_B TL_c/(4PU_0) $.
Thus, a dimensionless potential is obtained, which reads
\begin{equation}
U(x)= \mu\!\left\{\left[1-e^{-\beta(x-\rho)}\right]^2-1
 +A\!\left(\frac{1}{1-x}-1-x+2x^2\right) \right\},    \label{eq:berk-dimless}
\end{equation}
Note that, in order not to clutter our formulae, we have not
introduced a different notation for the dimensionless potential. The
values of the parameters therein are $\mu=0.0333$, $\beta=30$,
$\rho=0.133$ and $A=0.776$. In dimensionless variables, $F_{b}=0.527$
($52.7$pN) and $\ell_{b}=0.157$ ($4.70$nm). The relevant time scale
is set by the friction coefficient $\gamma$, $[t]= \gamma L_c/[F]$. In
turn, $\gamma$ is given by the Einstein relation $D=k_B T/\gamma$,
where $D$ is the diffusion coefficient for tethered proteins in
solution. We consider a typical value of $D$, also taken from
\cite{ber10}, $D =1500$ nm$^2$/s, so that $\gamma=0.0028$pN nm$^{-1}$
s.

\begin{figure}
  \centering
  \includegraphics[width=0.7  \textwidth]{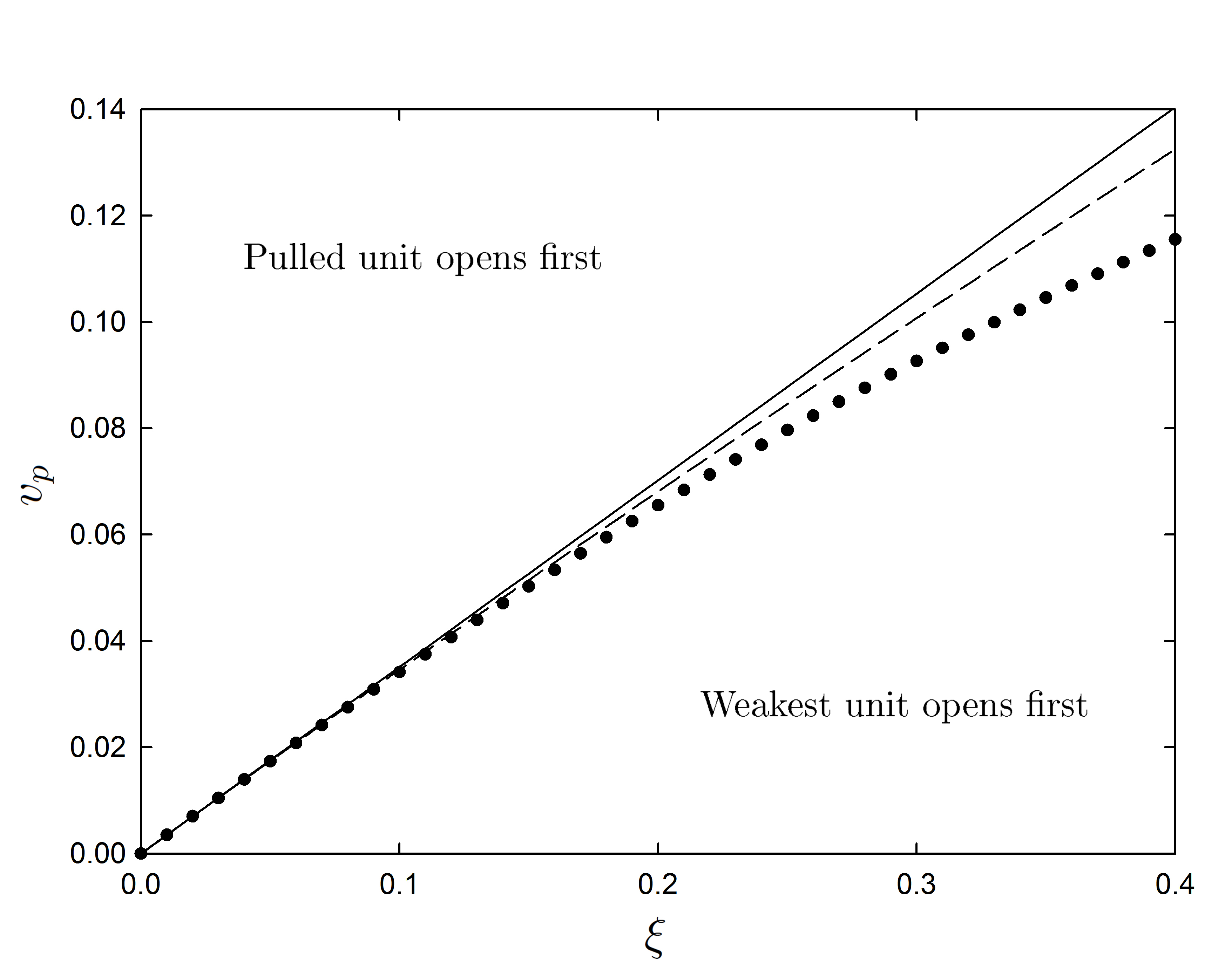}
  \caption{Phase diagram for the unfolding pathway in the pulling
    velocity-asymmetry plane for the Berkovich potential,
    (\ref{eq:berk-dimless}). Again, as in figure \ref{fig:5}, there
    appear two well-defined pulling regimes, separated by the curve
    giving the critical velocity $v_{c}$ as a function of the
    asymmetry $\xi$. The numerical values for $v_{c}$ (circles)
    compare very well with the theoretical expression (solid line),
    \eqref{eq:26a}. Again, the dashed line corresponds to the
    alternative approach discussed in the text for the quartic
    potential, which once more significantly improves the agreement
    theory-simulation for the larger values of $\xi$.  }
\label{fig:6}
\end{figure}

We consider a system of $4$ units, again with all the units but the
first being identical. Then, $U_{i}(x)=U(x)$, $i\neq 1$, and the first
unit is the weakest because $U_{1}(x)=(1-\xi)U(x)$. The situation is
then similar to the one we have already analyzed with the quartic
potential \eqref{eq:30}, but there is a difference that should be
noted: here, $U(x)$ is the free energy at zero force, whereas for the
quartic potential $U(x)$ was the free energy at a force $F_{0}$ for
which the folded and unfolded minima were equally deep. Then, the force
here must not be interpreted as the extra force from $F_{0}$, but as
the whole force that is applied to the polyprotein. On the basis of
our theory, we expect the simplest situation with only one critical
velocity $v_{c}$, below (above) which the weakest unit (the pulled
unit) unfolds first. This is also indeed the case in the numerical
simulations, and we compare the theoretical and numerical critical
velocities in figure \ref{fig:6}. A very good agreement is found
again, up to values of the asymmetry $\xi$ of the order $0.1-0.2$.

The above discussion shows that the validity of the theory presented
here is not restricted to simple potentials like the quartic one; on
the contrary, it can be confidently applied to experiments in which
the units are described by realistic potentials. Moreover, for the
typical parameters we are using, the theoretical critical velocity
$v_{c}$ for the Berkovich potential equals $1270$nm/s for an asymmetry
$\xi=0.1$, which can be regarded as quite a conservative estimate of
the largest asymmetries for which our theory gives an almost perfect
description of the unfolding pathway. Interestingly, this pulling
speed corresponds to the upper range of velocities usually employed in
AFM experiments, for instance see Table I of \cite{HyD12}. Therefore,
testing our theory in real AFM experiments with modular proteins
should be achievable.

Finally, we consider a more complex situation, in which more than one
unit is different from the rest and there may exist more than one
critical velocity. To be concrete, we have considered a system with
$4$ units in which $U_{2}(x)=U_{3}(x)=U(x)$, $U_{1}(x)=(1-\xi)U(x)$ as
before but $U_{4}(x)=(1+3\xi/2)U(x)$, with $U(x)$ being the quartic
potential in \eqref{eq:30}. In this situation, we have two different
critical velocities: for very low pulling speeds, the weakest unit is
the first to unfold, but there appears a velocity window inside which
neither the weakest nor the pulled unit is the first to unfold. This
stems from the fact that the first and the third unit reach
simultaneously the limit of stability for a velocity
$v^{1}(3)=4\xi \gamma^{-1} F_{b}/3$ that is smaller than the velocity
$v^{1}(4)=5\xi \gamma^{-1}F_{b}/3$ for which the first and the last
would do so. The physical reason behind this is the threshold force of
the pulled unit being larger enough than that of the third one.  We
recall that $v^{i}(j)$ is the velocity for which the $i$-th and the
$j$-th unit reach simultaneously their limits of
stability. Afterwards, the third unit and the fourth attain the limit
of stability in unison for a velocity $v^{3}(4)=2\xi\gamma^{-1}F_{b}$,
and the following picture emerges from our theory. Using the notation
introduced in section \ref{sec:critical_speed}, we define two critical
velocities,
  \begin{equation}
    \label{eq:1}
    \frac{\gamma v_{c}^{(1)}}{\xi}=\frac{4 F_{b}}{3}, \qquad
    \frac{\gamma v_{c}^{(2)}}{\xi}=2 F_{b},
  \end{equation}
  such that for $v_{p}<v_{c}^{(1)}$, it is the weakest unit that
    unfolds first, for $v_{c}^{(1)}<v_{p}<v_{c}^{(2)}$, it is the
    third unit that unfolds first, and, finally, for
    $v_{p}>v_{c}^{(2)}$, the first unit to unfold is the pulled one.

\begin{figure}
  \centering
  \includegraphics[width=0.7\textwidth]{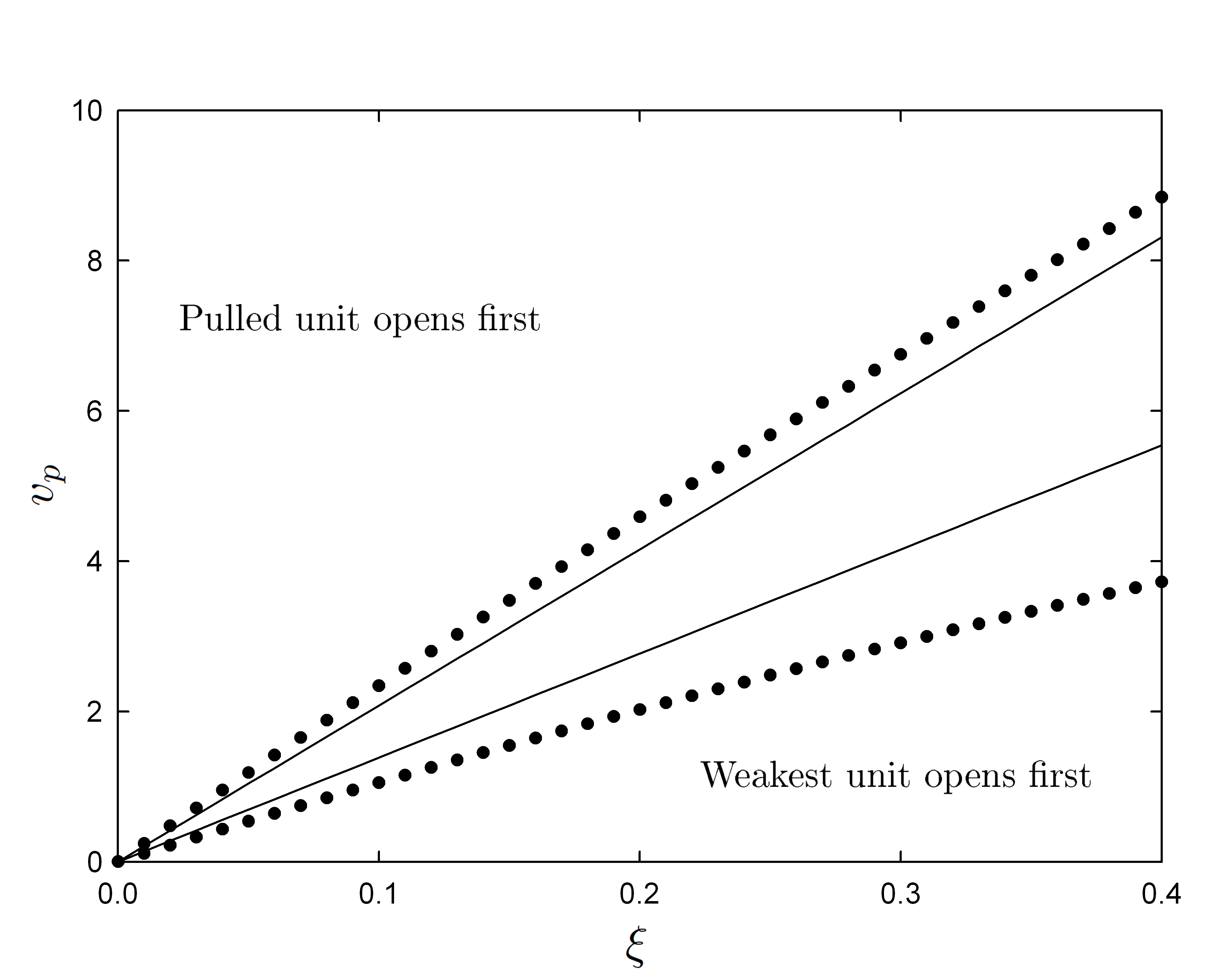}
  \caption{Phase diagram for the unfolding pathway in the pulling
    velocity-asymmetry plane for the more complex case discussed in
    the text. Now, we have three well-defined pulling regimes,
    separated by two curves giving the critical velocities
    $v_{c}^{(1)}$ and $v_{c}^{(2)}$ as a function of the asymmetry
    $\xi$. In this case, our theoretical approach is able to reproduce
    the existence of the three different pulling regimes, but the
    discrepancies between the theoretical and the numerical values for
    the critical velocities are larger than in the simpler cases
    considered in the previous figures.}
\label{fig:7}
\end{figure}

    We check the more complex scenario described in the previous
    paragraph in figure \ref{fig:7}. Therein, we observe that (i) our
    theory correctly predicts the existence of the three pulling
    regimes described above but (ii) even for very small asymmetries,
    there appear some noticeable discrepancy between theory and
    simulation. Since the validity of the perturbative expansion for
    obtaining the critical velocities from the condition
    (\ref{eq:stab}) is strongly supported by the accurateness of the
    theoretical prediction for the simplest case, see figures
    \ref{fig:5} and \ref{fig:6}, this discrepancy should stem from the
    intrinsically approximate character of the condition $U''_{i}=0$
    for determining the stability threshold when $v_{p}\neq 0$.
    Therefore, an improvement of the present theory should involve the
    derivation of a more accurate condition for obtaining the
    stability threshold in the case of finite pulling velocity. This
    point, which probably makes a multiple scale analysis necessary
    for lengths close to the condition $U''=0$, certainly deserves
    further investigation.

\section{\label{sec:conclusions}Conclusions}
We have investigated the general properties of the unfolding pathway
in pulled proteins by means of a simple model portraying the chain as
a sequence of nonlinear modules.  The deterministic approximation of
the Langevin equations controlling the time evolution of the units
extensions is used therein. This is a sensible approach in our
  model system: note that the free energies characterising the
  different units are considered to be very similar, and therefore
  their Kramers rates for thermal activated unfolding would also be
  very close. Then, were thermal effects important, our unfolding
  trajectories would be essentially stochastic and we would not
  observe a specific unfolding pathway.

Nevertheless, in some recent optical tweezers experiments in
  which thermal fluctuations are relevant, definite pathways have also
  been observed. Although trajectories are indeed stochastic, the
  pathway is well-characterised \cite{NYFWyW11,SZGGyR11}. Therefore,
  the analysis of these experiments needs a more sophisticated theory,
  which takes into account thermal noise effects. Moreover, it seems
  that there are other elements that should be incorporated, such as
  (i) the possible coupling between the different units and (ii) their
  dissimilar free energies. For example, the former may explain the
  existence of \textit{dead ends} observed in~\cite{SZGGyR11}, that
  is, intermediate states that do not allow the system to completely
  relax, whereas the latter may lead to Kramers rates leading to the
  separation of the timescales for the different unfolding events.

The equilibrium extensions of the units are governed by different free
energies, which we have called asymmetry or disorder in the
free energies. Our theory, based on stability considerations, is able
to explain the experimental observations: (i) for low pulling speeds,
it is the weakest unit that unfolds first, (ii) for large enough
pulling speed, it is the pulled unit that opens first.  This has been
done by introducing a perturbative expansion both in the asymmetry of
the free energies and in the pulling speed.  Moreover, our approach
makes it possible to identify a critical rate that separates two
well-defined regimes.  To the lowest order, this critical velocity has
a linear dependence on the asymmetry of the potential.  In spite of
the crude approximations, our theory compares quite well with the
numerical data even beyond the applicability regime.  Moreover, our
results provide a guide to interpret some inversions observed in the
sequence of unfolding of the stable regions of the maltose-binding
protein during its mechanical denaturation \cite{GMTCyC14,
aggarwal2011ligand}.

It must be stressed that to the lowest order in our theory, the system
is sweeping the stationary branches of the force-extension curve. In
this sense, the pulling process is very slow or
\textit{adiabatic}. Despite this adiabatic nature of the pulling
process, it is not always the weakest unit to unfold first.
As long as the pulling speed $v_{p}\neq 0$, the closer to the
pulled terminal one unit is, the larger the force acting on it. This
gradient in the distribution of the force across the protein,
which increases with the pulling speed, makes it possible that the
last unit reaches first its limit of stability.

We have limited ourselves to the investigation of the first unfolding
event. However, our argument can be easily generalized to the
next unfolding event: the difference is that the zero-th order
approximation is no longer given by all the units sweeping the
all-units-folded branch but by the sweeping of the branch with one
module unfolded and the remainder folded. Then, a similar perturbative
expansion around this zero-th order solution in powers of the
asymmetry and the pulling speed would give the next unit that opens.

If the biomolecule comprises several perfectly identical units,
  the asymmetry correction vanishes, because $\delta U_{i}=0$ (and thus
  $\delta f_{i}=0$) for all the units. In that case, our theory predicts
  that it is always the pulled unit that unfolds first. However, even
  in engineered modular proteins, slight differences from module to
  module may be present. In fact, this has lead to the analysis of the
  impact of quenched disorder in the force-extension curves of
  biomolecules \cite{BCyP15}. In the present context, we may also
  introduce stochastic free energy deviations, following a certain
  probability distribution. Next, our theoretical approach can be
  applied to this system with quenched disorder in the free energies.
  Interestingly, evidence of dynamical disorder, that is, a
  fluctuating environment, has been recently brought to bear in
  stretching experiments \cite{HHyT14}. The analysis of this situation
  needs a more complex theory, in which the free energy landscape
  fluctuates in time, and is outside of the scope of the present
  paper.

There are methods that extract the free energy landscape from
experimental data of pulling experiments, even when there are
  intermediates \cite{BCFyI04,LGyV08,AMJyR12,HGRyT13,MLyW15}. The
  resulting free energy is usually calculated as a function of the
  end-to-end distance of the molecule.  Nevertheless, in order to
  apply our theory, we do not need this global energy landscape as a
  function of the end-to-distance of the molecule but each unit's
  contribution thereto as a function of its own extension.  In
this regard, it is relevant to note that a similar velocity dependent
unfolding pathway should also be found in modular proteins, although
it has not been experimentally investigated to the best of our
knowledge.  In fact, we have analyzed a simple polyprotein model with
a realistic potential, and observed a completely analogous behaviour.
When all the modules are not identical, the weakest one
will always open first for small enough pulling velocities. On
  the other hand, if the pulled unit is not the weakest, this will no
longer be the case as the pulling speed is increased.  Since the free
energy of each module is experimentally accessible and the critical
velocity lies on the experimental range, a reliable test of our theory
could be done in modular proteins.  Another possibility that
  deserves attention is to test our theory in ankyrin repeat
  proteins. In \cite{LZZByM10}, the unfolding of a consensus ankyrin
  repeat protein, NI6C, has been investigated. This protein is
  composed of eight repeats: the two capping repeats are different
  from the six identical central ones, that at the C-terminus
  (N-terminus) is weaker (stronger) than the rest. Pulling from the
  C-terminus (weakest unit), Lee et al.~observe that the unfolding
  always starts from this end. This is what is expected from our
  theory, since when the pulled and the weakest units coincide, there
  is no competition between the asymmetry and kinetic terms.
  Therefore, it seems relevant to carry out the same experiment but
  pulling from the N-terminus (strongest unit), in which a much richer
  phenomenology is to be expected.

Very recently, sequential unfolding has been reported in a simple
model \cite{BCyP14}, which makes it possible to understand the
stepwise unfolding observed in force-clamp experiments with modular
proteins \cite{FyL04,WGDBByF07,LHMVyB13}. This sequential unfolding
appeared as a consequence of a depinning transition near the stability
threshold introduced by the coupling between nearest neighbour
units. Interestingly, the unfolding of the ankyrin repeat protein
  in \cite{LZZByM10} does not only start from a well-defined end
  but it is also sequential, which may hint at the significance of
  this kind of short-ranged couplings in the
  experiment.  Then, it seems also relevant to analyze whether a
similar sequential unfolding is present in the model developed
here, when a similar short ranged interaction between neighbouring
units is considered.

\ack
This work has been supported by the Spanish Ministerio de
Econom\'ia y Competitividad grant FIS2011-24460 (AP). CAP acknowledges
support by a PhD fellowship from Fundaci\'on C\'amara (Universidad de Sevilla).

\appendix
\section{\label{apa}Stability threshold}

To first order in $\xi$,  the extension $x_{i,b}$ such that
$U''_{i}(x_{i,b})=0$ verifies
\begin{equation}
  \label{eq:a0}
  U'''(\ell_{b})(x_{i,b}-\ell_{b})+\xi \deltaf'_{i}(\ell_{b})=0,
\end{equation}
that is,
\begin{equation}
  \label{eq:a1}
  x_{i,b}=\ell_{b}-\xi \frac{\deltaf'_{i}(\ell_{b})}{U'''(\ell_{b})}.
\end{equation}
The corresponding force at the stability threshold is obtained from
\eqref{eq:10}. To the lowest order in the deviations,
\begin{equation}
  \label{eq:28}
  F_{i,b}\equiv U'_{i}(x_{i,b})\sim U'_{i}(\ell_{b})=F_{b}+\xi\, \deltaf_{i}(\ell_{b}),
\end{equation}
because the next term, $U'''(\ell_{b})(x_{i,b}-\ell_{b})^{2}/2$, is of
the order of $\xi^{2}$. Therefore, the $i$-th module reaches its limit
of stability at the time for which $x_{i}=x_{i}^{(0)}+\xi\delta
x_{i}=x_{i,b}$, that is, when the length per unit $\ell$ has the value
$\ell_{i}$ verifying
\begin{equation}
  \label{eq:a2}
  \ell_{i}+\xi
\frac{\overline{\deltaf}(\ell_{i})-\deltaf_{i}(\ell_{i})}{U''(\ell_{i})}=
\ell_{b}-\xi \frac{\deltaf'_{i}(\ell_{b})}{U'''(\ell_{b})},
\end{equation}
or, equivalently,
\begin{equation}
  \label{eq:a3}
  \ell_{i}-\ell_{b}=\xi \frac{\deltaf_{i}(\ell_{i})-\overline{\deltaf}(\ell_{i})}{U''(\ell_{i})}-\xi \frac{\deltaf'_{i}(\ell_{b})}{U'''(\ell_{b})}.
\end{equation}
We know that $\ell\to\ell_{b}$ when $\xi\to 0$. But $U''(\ell_{b})=0$
and thus we cannot substitute $\ell_{b}$ on the rhs of \eqref{eq:a3}.
On the other hand, this means that the dominant balance for $\xi\to 0$ involves the
lhs and the first term on the rhs of \eqref{eq:a3}.
Therefore, making use of $U''(\ell)\sim
U'''(\ell_{b})(\ell-\ell_{b})$, we get
\begin{equation}
  \label{eq:a4}
  (\ell_{i}-\ell_{b})^{2} \sim \xi \frac{\deltaf_{i}(\ell_{b})-\overline{\deltaf}(\ell_{b})}{U'''(\ell_{b})}.
\end{equation}
Since $U'''(\ell_{b})<0$ (see figure \ref{fig:2}), this means that only
the units with $\deltaf_{i}(\ell_{b})$ smaller than the average (that
is, weaker than average) reach the limit of stability in the limit as
$v_p \to 0$.
In fact, it is the weakest unit, that is, the unit with
smallest $\deltaf_{i}(\ell_{b})$, that unfolds first.

It is interesting to note that, in order to obtain \eqref{eq:a4},
we have completely neglected the last term on the rhs of
\eqref{eq:a3}.
Since, in turn, this term stems from the last term
on the rhs of \eqref{eq:a1}, to the lowest order we are solving
the equation $x_{i}=\ell_{b}$. In other words, to the lowest order the
stability threshold can be considered to be given by the
non-disordered, zero-asymmetry case, free energy $U(x)$. For the
sake of concreteness and simplicity, we have stuck to the asymmetry
contribution $\delta x_{i}$ in this appendix, but the same condition
$x_{i}=\ell_{b}$ would still be valid, had we taken into account the
kinetic contribution $\Delta x_{i}$ derived in
section \ref{sec:pulling_speed}. The reason is that there is also a
factor $U''(\ell)$ in the denominator of $\Delta x_{i}$, see
\eqref{eq:22a}, and thus both the terms coming from $\delta x_{i}$ and
$\Delta x_{i}$ are dominant against the last term on the rhs of
\eqref{eq:a1}.

 \section{\label{ap1}Discrete inhomogeneous diffusion equation}
 In this appendix, we briefly discuss a general procedure which is
 useful to solve linear difference equations similar to those in
 \eqref{eq:18} and \eqref{eq:21}. The methods for solving
 difference equations often resemble those used for solving
 analogous differential equations; the latter may be thought of as the
 continuous limit of the former. Both  \eqref{eq:18} and \eqref{eq:21}
 belong to the following general class of second-order linear
 difference equations for $y_{i}$,
\begin{subequations}\label{eq:B1}
\begin{eqnarray}
&y&_{1}= g(y_{2},\ldots , y_{N-1}), \label{eq:B1a} \\
\label{eq:B1b} &y&_{i+1}+y_{i-1}-2 y_{i}  =  K, \quad 1<i<N, \\
&y&_{N} = h(y_{2},\ldots , y_{N-1}), \label{eq:B1c}
\end{eqnarray}
\end{subequations}
in which $g$, $h$ are arbitrary functions and $K$ is a given
constant. (\ref{eq:B1b}) is a second-order linear difference
equation, and (\ref{eq:B1a}) and (\ref{eq:B1c}) are its boundary
conditions. It may be thought of as a discrete inhomogeneous diffusion
equation: $y_{i+1}-y_{i}$ is the first discrete derivative, so that
$y_{i+1}+y_{i-1}-2y_{i}$ is the second discrete derivative
\cite{ByO99}. In complete analogy with the corresponding differential
equation $y''=K$, the general solution of (\ref{eq:B1b}) is
\begin{equation}\label{eq:B3}
y_i=c_{0}+c_{1}i+\frac{K}{2} i^{2},
\end{equation}
in which $c_{0}$ and $c_{1}$ are two arbitrary constants. The solution
of (\ref{eq:B1}) is, as usual, univocally determined by the boundary
conditions, from which specific values for $c_{0}$ and $c_{1}$ are
obtained.

Let us show that both \eqref{eq:18} and
\eqref{eq:21} can be cast in the above form. Firstly, in
\eqref{eq:18b}, it is easily seen that if we define
$y_{i}=\delta x_{i}+\delta f_{i}/U''(\ell)$, (\ref{eq:B1b}) is
obtained with $K=0$. Secondly, in \eqref{eq:21b}, it is
straightforward to identify $y_{i}=\Delta x_{i}$ and
$K=\gamma[NU''(\ell)]^{-1}$. A
simple calculation gives the constants $c_{0}$ and $c_{1}$ in
\eqref{eq:B3} for each case,  and thus the expressions for $\delta
x_{i}$ and $\Delta x_{i}$ in the main text.

\section{\label{apb}Order of the critical velocities}
Here, we prove that $v_c^{(k+1)}>v_c^{(k)}$. It is easy to show that
this inequality follows if we have that
\begin{equation}
  \label{eq:29}
  \deltaf_{\alpha_{k+2}}(\ell_{b})>\frac{\deltaf_{\alpha_{k+1}}(\ell_{b})
    (\nu_{k+2}-\nu_{k}) -\deltaf_{\alpha_{k}}(\ell_{b})
    (\nu_{k+2}-\nu_{k+1})}
    {\nu_{k+1}-\nu_{k}},
\end{equation}
in which $\nu_k=\alpha_{k}(\alpha_{k}-1)$. Due to \eqref{eq:vck}, $\alpha_{k+1}$ minimize
$v^{\alpha_{k}}(j)$. Therefore, in particular,
$v^{\alpha_{k}}(\alpha_{k+1})<v^{\alpha_{k}}(\alpha_{k+2})$, which is readily
shown to be equivalent to \eqref{eq:29}.

\newpage

\end{document}